\documentclass[twocolumn, twocolappendix]{aastex631}

\usepackage{graphicx}	% Including figure files
\usepackage{amsmath}	% Advanced maths commands
\usepackage{ulem}
\usepackage[T1]{fontenc}
\usepackage{comment}
\usepackage{orcidlink}
\usepackage{newtxtext,newtxmath}
\usepackage{xcolor}
\usepackage{CJK}
\usepackage{xpatch}
\usepackage{hyperref}

\hypersetup{
	colorlinks=true,
	breaklinks=true,
	citecolor=blue,
	allcolors=blue,
	frenchlinks=true
}

\makeatletter
\xpatchcmd\NAT@citex
{%
	\@citea\NAT@hyper@{%
		\NAT@nmfmt{\NAT@nm}%
		\hyper@natlinkbreak{\NAT@aysep\NAT@spacechar}{\@citeb\@extra@b@citeb}%
		\NAT@date
	}%
}
{%
	\@citea
	\NAT@nmfmt{\NAT@nm}%
	\NAT@aysep\NAT@spacechar
	\NAT@hyper@{\NAT@date}%
}
{}{}
\xpatchcmd\NAT@citex
{%
	\@citea\NAT@hyper@{%
		\NAT@nmfmt{\NAT@nm}%
		\hyper@natlinkbreak{\NAT@spacechar\NAT@@open\if*#1*\else#1\NAT@spacechar\fi}%
		{\@citeb\@extra@b@citeb}%
		\NAT@date
	}%
}
{
	\@citea
	\NAT@nmfmt{\NAT@nm}%
	\NAT@spacechar\NAT@@open\if*#1*\else#1\NAT@spacechar\fi
	\NAT@hyper@{\NAT@date}%
}
{}{}
\makeatother

\begin{document}
\begin{CJK*}{UTF8}{bkai}

\title{Simulating Radio Synchrotron Morphology, Spectra, and Polarization of Cosmic Ray Driven Galactic Winds}

\author[0000-0002-7401-382X]{H.-H. Sandy Chiu}
\affiliation{Department of Astronomy, University of Michigan, Ann Arbor, MI 48109, USA}

\author[0009-0002-2669-9908]{Mateusz Ruszkowski}
\affiliation{Department of Astronomy, University of Michigan, Ann Arbor, MI 48109, USA}
\affiliation{Max-Planck-Institut f{\"u}r Astrophysik (MPA), Karl-Schwarzschild-Str. 1, 85748 Garching, Germany}
\email{E-mail: mateuszr@umich.edu (MR)}

\author[0000-0002-7443-8377]{Timon Thomas}
\affiliation{Leibniz-Institute for Astrophysics Potsdam (AIP), An der Sternwarte 16, 14482 Potsdam, Germany}

\author[0000-0003-4984-4389]{Maria Werhahn}
\affiliation{Max-Planck-Institut f{\"u}r Astrophysik (MPA), Karl-Schwarzschild-Str. 1, 85748 Garching, Germany}

\author[0000-0002-7275-3998]{Christoph Pfrommer}
\affiliation{Leibniz-Institute for Astrophysics Potsdam (AIP), An der Sternwarte 16, 14482 Potsdam, Germany}
%\linenumbers

\newcommand{\dd}{\mathrm{d}}
\newcommand{\cmr}[1]{{\color{red}\bf{[MR: #1]}}}
\newcommand{\mr}[1]{{\color{red}#1}}
\newcommand{\ccp}[1]{{\color{orange}\bf{[CP: #1]}}}
\newcommand{\hhc}[1]{{\color{blue}\bf{[SC: #1]}}}
\newcommand{\mw}[1]{{\color{blue}#1}}
\newcommand{\rv}[1]{\color{violet}#1}


\begin{abstract}
The formation of galaxies is significantly influenced by galactic winds, possibly driven by cosmic rays due to their long cooling times and better coupling to plasma compared to radiation. In this study, we compare the radio observations of the edge-on galaxy NGC 4217 from the CHANG-ES collaboration catalog with a mock observation of an isolated galaxy based on the \textsc{arepo} simulation that adopts the state-of-the-art two-moment cosmic ray transport treatment and multiphase interstellar medium model. We find significant agreement between the simulated and observed images and spectroscopic data for reasonable model parameters. Specifically, we find that (i) the shape of the intensity profiles depends weakly on the magnitude of the magnetic field, the distance of the simulated galaxy, and the normalization of the CR electron spectrum. The agreement between the mock and actual observations is degenerate with respect to these factors; (ii) the multi-wavelength spectrum above 0.1~GHz is in agreement with the radio observations and its slope is also only weakly sensitive to the magnetic field strength; (iii) the magnetic field direction exhibits X-shaped morphology, often seen in edge-on galaxies, which is consistent with the observations and indicates the presence of a galactic-scale outflow. Our results highlight the importance of incorporating advanced cosmic ray transport models in simulations and provide a deeper understanding of galactic wind dynamics and its impact on galaxy evolution. 
\end{abstract}

\keywords{cosmic rays -- galaxies: magnetic fields -- radio continuum: galaxies -- methods: numerical -- radiation mechanisms: non-thermal} %https://journals.aas.org/keywords-2013/

\section{Introduction} \label{sec:intro}
Most galaxies exhibit baryon fractions significantly below the average cosmological value. 
The missing baryons are either not captured by the developing protogalaxies' potential wells or are expelled due to feedback mechanisms during the galaxy formation process \citep{anderson_hot_2010, somerville_physical_2015}.
In halos with masses comparable to that of the Milky Way, which have the most effective star formation efficiency, only around 20\% of the available baryons are converted into stars \citep{bregman_search_2007, moster_constraints_2010, tumlinson_circumgalactic_2017}. 
In galaxies exceeding the mass of the Milky Way, active galactic nuclei are believed to play a significant role in supplying the necessary feedback to suppress excessive star formation \citep{croton_many_2006}, while in less massive galaxies, starburst activities are widely accepted to be the main mechanism of generating outflows and thus limit the amount of gas mass available for star formation
\citep{veilleux_galactic_2005, bland-hawthorn_galactic_2007}.\\
\indent
In the traditional framework of galactic winds powered by supernovae \citep{chevalier_wind_1985}, galactic outflows are initiated by thermal energy, propelling the gas out of the galaxy ballistically. Although these models have demonstrated the ability to generate superwinds observed in starburst galaxies \citep{bustard_versatile_2016}, they are still facing obstacles such as the overcooling problem \citep{walch_silcc_2015, naab_theoretical_2017}, where the thermal energy released by supernovae is rapidly dissipated, leading to a decrease in the effectiveness of supernova explosions in regulating star formation. Feedback from supernovae can also be driven by radiation pressure \citep{murray_maximum_2005, murray_radiation_2011, hopkins_stellar_2012}. However, this model also faces challenges in effectively coupling stellar radiation to the gas \citep{rosdahl_galaxies_2015}, resulting in reduced momentum transferred to the gas and allowing radiation to escape through a path of least resistance. 
Furthermore, reprocessing of UV and optical radiation from massive stars is not expected to have a significant dynamical effect on the surrounding gas due to the reprocessing of this radiation to far-IR, where the coupling of radiation to the gas is weak \citep{reissl2018,menon2022}.
These processes reduce the efficiency of stellar feedback and make both the thermal energy-driven and radiation-pressure-driven models insufficient in explaining observations \citep{steidel_structure_2010}.
\\
\indent
Cosmic rays (CRs) have recently gained significant attention for their role in galactic feedback \citep{ruszkowski_cosmic_2023}. These relativistic particles accelerated in supernova remnants (SNRs) \citep{blandford_particle_1987, caprioli_2014} or by winds of massive young stars \citep{bykov_nonthermal_2014}, are relatively sparse in the interstellar medium (ISM). However, observations from galaxies with distinct outflow characteristics such as M82 and NGC 253 \citep{ackermann_gev_2012, bolatto_suppression_2013, yoast-hull_winds_2013} suggest that the CR energy density is comparable to, or even exceeds, that of thermal gas and magnetic fields \citep{acciari_connection_2009, paglione_properties_2012}. These observations indicate that CRs can drive, shape, and alter large-scale galactic outflows.\\
\indent
In contrast to traditional feedback models, where thermal energy is injected at SNR sites and quickly radiated away, injection of a fraction of SN energy in the form of CRs can result in a prolonged dynamical impact of CRs on the gas due to relatively long CR cooling times. Thus, CR feedback offers a possible solution to the ``overcooling" problem \citep{jubelgas_cosmic_2008}. Following their injection at the SNR sites, CRs can be transported via diffusion, advection, and streaming to a low-density region above the galactic disk, where they can accelerate this gas more effectively than radiation due to better coupling of CRs to the gas compared to radiation \citep{socrates_eddington_2008, huang_launching_2022}. Consequently, CRs can be a critical feedback agent, the efficiency of which depends sensitively on the properties of CR transport.\\
\indent
This motivates the need to constrain CR transport processes.
Edge-on galaxies serve as good laboratories for studying CR transport models because in these objects non-thermal emission generated by CRs can be directly observed above and below galactic disks. One-dimensional CR transport models \citep{heesen_advective_2016, heesen_radio_2018} have been developed to constrain transport processes and quantify the effective CR diffusion coefficient. However, these models rely on simplifying assumptions regarding the magnetic field and suffer from uncertainties associated with the modeling of thermal emission. While one-dimensional models are valuable tools for constraining physical parameters and understanding individual observations, their reliance on these assumptions restricts the predictive capability of this approach.\\
\indent
Magneto-hydrodynamical simulations of supernova feedback including simplified prescriptions for CR transport (i.e., spatially-constant diffusion coefficient or constant streaming velocity) have demonstrated that CR pressure forces can drive galactic winds \citep[e.g.,][]{Breitschwerdt1991, uhlig_2012, Salem2014, pakmor_2016}. Models that include this physics can reproduce electron and proton spectra in the local environment observed by Voyager-I and AMS-02 \citep{werhahn_cosmic_2021a}, FIR--$\gamma$-ray correlation \citep{pfrommer_2017,werhahn_cosmic_2021b}, and the FIR-radio correlation \citep{werhahn_cosmic_2021c,pfrommer_2022}.
However, in a realistic situation, CR transport is likely to include a combination of diffusion and streaming depending on the CR scattering frequency with low and high scattering rates corresponding to diffusion and streaming, respectively. Simulations that include these effects have been performed \citep[e.g.,][]{ruszkowski_global_2017,wiener_cosmic_2017}, but have not faced the same level of scrutiny in the context of specific observational predictions and comparisons to the observational data as those based on the above simplified CR transport prescriptions. 
Furthermore, CR transport is generally expected to be highly spatially variable
\citep[e.g.,][]{farber_impact_2018,Hopkins2021}. By comparing the streaming instability growth rate to the dominant wave damping rates -- ion neutral and non-linear Landau damping rate -- one can infer the relative contributions to CR flux due to diffusion and streaming \citep{Armillotta2021,Armillotta2022}. The resulting effective CR transport speeds are very large in the predominantly neutral ISM, which imposes severe limitations on the numerical scheme requiring prohibitively small timesteps for numerical stability.\\
\indent 
The above considerations motivate development of more efficient computational methods that can tackle more realistic spatially variable and fast CR transport without sacrificing numerical accuracy, and thus enable more meaningful comparisons to observations. 
This challenge can be successfully addressed by employing two-moment methods, which draw on the analogy between CR and radiative transport and extend the set of governing equations to include explicit time evolution of the CR flux rather than assuming that it is given by its steady-state form \citep{jiang_new_2018}. This treatment can be further extended to include explicitly evolving the energy density in the Alfv{\'e}n waves responsible for scattering CRs \citep{thomas_cosmic-ray_2019, thomas_cosmic-ray-driven_2023}. 
The two-moment method has been tested against observations of radio harps near the Galactic center \citep{heywood_inflation_2019}. These radio harps are likely created by CRs emitted by a moving source such as a pulsar wind nebula or stellar wind and are injected into the ambient magnetic field. These CRs follow magnetic field lines and encode their transport history. 
Comparisons of the observed brightness profiles to those simulated using the two-moment method, demonstrate that CR transport is best described by a combination of streaming and diffusion \citep{thomas_probing_2020}.\\
\indent
Given the successful validation of the two-moment method on small scales, this study aims to determine if this method, when applied to CR transport governed by the streaming instability, diffusion in the presence of both the non-linear Landau and ion-neutral damping, and state-of-the-art multiphase ISM model, can produce galactic outflow properties that agree with the observational data. To this end, we generate mock observations from a state-of-the-art galactic wind simulation \citep{thomas_2024arXiv} and compare them with the observational data from an edge-on galaxy NGC 4217 \citep{stein_chang-es_2020}. 
Our work complements that of \citet{ponnada_synchrotron_2024}, who presented comparisons of mock radio observations based on cosmological simulations of Milky Way-mass galaxies using a CR magneto-hydrodynamical approach with on-the-fly computation of the CR spectrum, but differs in several respects. Our two-moment transport method includes explicit evolution of the energy density in Alfv{\'e}n waves responsible for scattering CRs and includes a different approach to modeling the ISM physics. 
Our ISM modeling includes a thermochemistry model incorporating important ionization and low-temperature cooling and heating processes. This allows us to faithfully evolve the thermodynamic state of the diffuse ISM and also to model the effects of ion-neutral damping based on atomic collision data and its subsequent impact on CR transport. We also focus on different types of observables such as the spatially-resolved profiles of radio intensity, morphology of the polarization fraction and field orientation (including Faraday rotation), and radio spectrum. 
\\
\indent
This paper is structured as follows: In Section \ref{sec:method}, we present the simulation setup and describe how we calculate steady-state CR spectra, radio emission, and Stokes parameters. In Section \ref{sec:result}, we compare mock observations to observational data focusing on the intensity profiles, radio emission morphology, polarization, and multi-wavelength spectrum of the galaxy. Finally, we outline possible caveats and summarize key findings in Section \ref{sec:conclusion}. 
In Appendix~\ref{app:steady-state}, we examine the steady-state assumption of our approach of modeling CR electron spectra. 
In Appendix~\ref{app:rotation measure} we present the Faraday rotation measure of our galaxy. 
In Appendix~\ref{app: x-shape field} we show how the magnetic field in the disk is effected by the foreground and background gas. 
Finally, we show the vertical synchrotron profiles in five strips in Appendix~\ref{app:strip}, which are lined up horizontally along the disk and probe different outflow regions of the galaxy ranging from the outskirts to the nuclear region.

\section{Methods} \label{sec:method}
In this section, we describe the simulation used in this study. This is followed by the description of our method to calculate via post-processing the CR spectra, radio emission, and polarized intensity.

\subsection{Simulation}
The simulation used in this study was performed using the moving-mesh code \textsc{arepo} \citep{springel_e_2010, pakmor_improving_2016} with its two-moment CR magnetohydrodynamics module \citep[CRMHD,][]{thomas_cosmic-ray_2019, thomas_finite_2021, thomas_comparing_2022} and the \textsc{crisp} ISM physics module (Thomas et al.,  \textcolor{blue}{in prep.}, see also \citealt{thomas_2024arXiv}). Below, we first describe our simulation settings, then provide an overview of the two-moment CR method and the \textsc{crisp} framework.\\
\indent
We study an isolated Milky Way-type galaxy with a dark matter halo mass of $M_{\rm 200}=10^{12}~\mathrm{M}_\odot$. The simulation starts from an idealized setup that includes a doubly exponential disk for both the gaseous and stellar components. The gaseous disk has a radial scale length of 5~kpc, a scale height of 0.5~kpc, and a total mass of $8\times 10^9~\mathrm{M}_\odot$. The stellar disk has the same scale length and height, but a total stellar mass of $3.2\times 10^{10}~\mathrm{M}_\odot$. Star particles are initialized with individual particle masses of $M_*=1000~\mathrm{M}_\odot$ placed randomly following the exponential disk profile. 
Gas particles are initialized in the same way and are (de-)refined based on the more stringent of two criteria: (i) the mass of the gas particle should differ by a factor of at most 2 (at least 0.5) from $1000~\mathrm{M}_\odot$ and (ii) the volume of the gas particle should differ by a factor of at most 2 (at least 0.5) from $V = (4\pi/3)r^3$ where the effective resolution is better than or equal to $r=100$~pc in our region of interest. The latter criterion ensures that our numerical resolution is approximately an order of magnitude finer than the beam size of the observation.\\
\indent
Star formation is modeled following a Schmidt-type approach that converts gas in cells whose number density is above the threshold of 100 hydrogen particles per cm$^3$ into new stars within a free-fall time, which is 4.5~Myr at this threshold. These newborn stars participate in the stellar feedback process by ejecting mass, metals, and energy into the ISM, with respective rates based on \texttt{STARBURST99} calculations \citep{leitherer_starburst99_1999}. 
We assume that each supernova injects $1.06\times 10^{51}$~erg of mechanical energy and that 5\% of the mechanical energy supplied by SNe and stellar winds is injected as CRs. The CR energy density is then evolved using the two-moment CRMHD model of \citet{thomas_cosmic-ray_2019}. 
The initial magnetic field consists of a toroidal magnetic field with a strength of $B_{\rm tor}=10^{-1}\sqrt{\rho/\rho_{\rm max}}~\mu {\rm G}$, where $\rho$ is the gas density and $\rho_{\rm max}$ is the maximum gas density at the galaxy center, and a constant vertical magnetic field with a strength of $10^{-3}~\mu$G. The final magnetic field is unaffected by the initial conditions as the field strength is significantly amplified by the small-scale dynamo process until the system reaches equilibrium.\\
\indent
Our isolated galaxy simulation aims to test the two-moment CRMHD method and the multi-phase ISM model \textsc{crisp}. The two-moment CRMHD method \citep{thomas_cosmic-ray_2019} is used to evolve the CR energy density and CR flux density. The CR transport velocity is determined by how the CRs interact with small-scale Alfv\'en waves, which can be damped by non-linear Landau effect \citep{miller_magnetohydrodynamic_1991} or ion-neutral damping \citep{kulsrud_effect_1969}. In addition to the energy loss due to the interactions of CRs with Alfv\'en waves, CRs can also lose energy through hadronic and Coulomb interactions \citep{pfrommer_simulating_2017}. The CR energy density is evolved assuming the gray approach (i.e., we do not explicitly track CR energy density corresponding to different momenta of CR particles) and the CR energy loss rates are averaged over the spectrum of CR momentum.\\
\indent
We use the \textsc{crisp} module \citep{thomas_2024arXiv} to incorporate the multi-phase physics of the ISM. \textsc{crisp} tracks the ionization stages of 12 different species, including $\rm H_2$, \ion{H}{1},  \ion{H}{2}, all ionization stages of He, and the first two ionization stages of C, O, and Si. The package evolves the ionization stages considering heating from the far ultraviolet (FUV) photons emitted by newly born stars and various cooling processes including Ly$\alpha$ cooling by H, rotation-vibrational lines from $\rm H_2$ \citep{moseley_turbulent_2021}, bremsstrahlung cooling at high temperatures \citep{cen_hydrodynamic_1992}, and metal line cooling calculated directly from collision rates \citep{abrahamsson_fine-structure_2007, grassi_krome_2014} at low temperatures ($T < 10^4$~K) or by interpolating tables pre-compiled using the \textsc{chianti} code \citep{dere_chianti_1997} at high temperatures ($T > 10^4$~K). 

\subsection{Modeling Cosmic-Ray Spectra} \label{sec:model spectra}
We followed the procedure described in \cite{werhahn_cosmic_2021a} to model CR spectra using the \textsc{Crayon+} code. As this approach adopted a spatially constant diffusion coefficient, here we outline the key steps in the calculation and discuss the modifications needed to accommodate spatially varying CR transport due to a combination of diffusion and streaming in the two-moment model.\\
\indent
In order to compute the energy distribution of CR particles, we consider a cell-based steady-state approximation, which assumes that the timescale over which the CR energy density changes is longer than the proton cooling timescale, which is generally longer than the electron cooling timescale. This assumption is validated in both \cite{werhahn_cosmic_2021a} and Appendix \ref{app:steady-state}. We then solve the diffusion-loss equation separately in each cell for protons, primary electrons, and secondary electrons. Under the steady-state assumption, the source and loss term in the diffusion-loss equation are in balance such that the total energy of the CR population in each cell remains constant in time.\\
\indent
We model the source term $q_i=q_i(p_i)$ as a power-law function of momentum with an exponential cutoff:
\begin{equation}
    q_i(p_i)\dd p_i=C_ip_i^{-\alpha_{\rm inj}}\exp{[-(p_i/p_{\rm cut,i})^n]}\dd p_i
    \label{eq:injection spectrum}
\end{equation}
where $p_i=P_i/(m_ic)=\sqrt{[E_i/(m_ic^2)]^2 - 1}$ is the normalized particle momentum and $E_i$ is the CR total particle energy, subscript $i$ specifies CR species ($= {\rm p, e}$ for protons and electrons, respectively), $m_i$ is the proton/electron rest mass, and $c$ is the speed of light. The normalization of the source function $C_i$ is determined through subsequent re-normalization processes (see below).
We adopt the injection spectral index $\alpha_{\rm inj}=2.1$ for both protons and primary electrons, cut-off momentum of protons $p_{\rm cut, p}=1\ {\rm PeV}/m_{\rm p}c^2$ \citep{gaisser_cosmic_1990} and electrons $p_{\rm cut, e}=20\ {\rm TeV}/m_{\rm e}c^2$ \citep{vink_supernova_2011}, and $n=1$ for protons and $n=2$ for primary electrons \citep{zirakashvili_analytical_2007, blasi_shock_2010}.
The source function of secondary electrons and positrons is calculated from the steady-state proton spectrum and the parametrization of the differential cross section of secondary particle production from \cite{Kelner_crosssection_2006} at large proton energies ($T_{\rm p}>100$~GeV). For $T_{\rm p}<10$~GeV, we use the parametrization from \cite{Yang_crosssection_2018} and interpolate between the two in the intermediate range \citep[see equations B1, B5 and B6 in ][]{werhahn_cosmic_2021a}.\\
\indent
The loss term in the diffusion-loss equation consists of cooling and escape losses. For the cooling losses, we adopt the same approach as that described in \cite{werhahn_cosmic_2021a}. Specifically, we include hadronic and Coulomb interactions for protons, as well as synchrotron and inverse Compton (IC) losses for primary electrons \citep[as detailed in sections 3.1.2–3.1.3 in][]{werhahn_cosmic_2021a}. 
In contrast to \cite{werhahn_cosmic_2021a} where only advection and diffusion escape losses were included, we include escape losses due to advection, diffusion, and streaming losses. In \cite{werhahn_cosmic_2021a}, the diffusion timescale is calculated by $\tau_{\rm diff} = L_{\rm CR}^2/D(E)$, where $E$ is total energy of the CR spectral bin, $L_{\rm CR}=e_{\rm CR}/|\nabla e_{\rm CR}|$ is the diffusion length in each cell, $e_{\rm CR}$ is the energy density of the CR fluid, and $D(E)=D_0\ (E/E_0)^{0.3}$ is the energy-dependent diffusion coefficient, where $D_0=10^{28}~{\rm cm^2 s^{-1}}$ and $E_0=3~{\rm GeV}$ are free parameters. 
This particular choice of a typical CR energy $E_0$ is motivated by the fact that CRs that contribute most to the pressure of the CR fluid have energies close to the peak in the CR spectrum that occurs at that characteristic energy \citep[e.g.,][]{ruszkowski_cosmic_2023}. 
Unlike in the simulations in \cite{werhahn_cosmic_2021a}, CR energy flux $f_{\rm CR}$ and the streaming velocity are explicitly modeled in this work. This allows us to define an effective diffusion coefficient as $D_{\rm 0, eff}=|\varv_{\rm CR}|L_{\rm CR}$, where the velocity of the CR fluid is given as 
\begin{equation}
    \varv_{\rm CR} = \frac{f_{\rm CR}}{e_{\rm CR}+P_{\rm CR}} = \frac{3f_{\rm CR}}{4e_{\rm CR}}
\end{equation}
where $P_{\rm CR}=(\gamma_{\rm CR}-1)e_{\rm CR}$ is the CR pressure and $\gamma_{\rm CR}=4/3$ \citep{thomas_cosmic-ray-driven_2023}.
With such-defined effective diffusion coefficient, the diffusion timescale (that indirectly includes streaming) becomes $\tau_{\rm diff}=L_{\rm CR}^2/D_{\rm eff}(E)=[L_{\rm CR}/|\varv_{\rm CR}|] (E/E_0)^{-0.3}$. 
The advection timescale is calculated by $\tau_{\rm adv}=L_{\rm CR}/\varv_z$, where $\varv_z$ is the velocity in the $z$-direction. In order to account for CR losses, we only include in the calculation the computational cells that include gas leaving the disk, i.e., whose for which $\varv_z>0$ for $z>0$ and $\varv_z<0$ for $z<0$ \citep[see][for a justification of the neglect of velocities in the disk]{werhahn_cosmic_2021a}.
Although one could in principle determine whether diffusion or advection is the dominant mechanism by comparing the corresponding timescales, preponderance of one mechanism does not imply that the other one is unnecessary. \cite{blasi_escape_2019} show that slow diffusion within the disk can generate a CR pressure gradient, thereby initiating CR outflow. Subsequently, advection transports the CRs to higher latitudes. This suggests that while advection may dominate in certain regimes, the initial conditions set by diffusion are critical for the overall dynamics of the system.

%%%%%% - Normalization of CR spectrum
\indent
After solving the diffusion-loss equation, we renormalize the steady-state spectra of CR protons such that the total kinetic energy density of the CR population $\int^\infty_0 T_{\rm p}(p_{\rm p})f_{\rm p,0}(p_{\rm p})\dd p_{\rm p}$ equals the CR energy density in our CRMHD model in each cell, where $T_{\rm p}(p_{\rm p})/(m_{\rm p}c^2)=\sqrt{p_{\rm p}^2+1}-1$ is the kinetic energy of each momentum bin and $f_{\rm p,0}$ is the CR proton spectrum.
In order to obtain the normalization of the CR electron spectrum, we assume a constant electron-to-proton injection number ratio $K^{\rm inj}_{\rm ep}\approx$ 0.092 at 10~GeV. Specifically, $K^{\rm inj}_{\rm ep}$ is defined as
\begin{equation}
    K^{\rm inj}_{\rm ep}(T_{\rm norm}) = \frac{q^{\rm prim}_{\rm e}(p_{\rm norm, e})}{q_{\rm p}(p_{\rm norm,p})}\frac{\dd p_{\rm e}}{\dd p_{\rm p}},
    \label{eq:kinj}
\end{equation}
where $T_{\rm norm}=10$~GeV is a fixed normalization kinetic energy for both electrons and protons, 
$q_{\rm p}$ and $q^{\rm prim}_{\rm e}$ are the injection spectrum for protons and primary electrons as defined in Equation~\eqref{eq:injection spectrum}, and $p_{\rm norm, e/p}=\sqrt{[T_{\rm norm}/(m_{\rm e/p} c^2)+1]^2 - 1}$. This shows that $K^{\rm inj}_{\rm ep}$ is independent of the kinetic energy of CRs as long as $T_{\rm norm}\gg m_{\rm e/p}c^2$ and the exponential cutoff term can be neglected in Equation~\eqref{eq:kinj}, i.e., $p_{\rm norm, i}\ll p_{\rm cut, i}$. The former criterion is approximately fulfilled for our chosen parameters and the associated small correction is discussed below (see the end of this section), while the latter criterion is 
very accurately met for our chosen parameters.
This value of $K^{\rm inj}_{\rm ep}$ can be further related to the electron-to-proton energy injection fraction 
\begin{equation}
    \zeta_{\rm prim}=\frac{\epsilon^{\rm inj}_{\rm e}}{\epsilon^{\rm inj}_{\rm p}}=\frac{\int^\infty_0 T_{\rm e}q_{\rm e}\dd p_{\rm e}}{\int^\infty_0 T_{\rm p}q_{\rm p}\dd p_{\rm p}}
\end{equation}
by
\begin{equation}
    K^{\rm inj}_{\rm ep}=\zeta_{\rm prim,no-cutoff}\left(\frac{m_{\rm p}}{m_{\rm e}}\right)^{2-\alpha_{\rm p}}
    \label{eq:kep}
\end{equation}
as long as the cut-off term can also be neglected in the calculation of $\epsilon^{\rm inj}_{\rm p/e}$. We find that this criterion is fulfilled as the integral values are very similar (10 percent difference for $\alpha_{\rm p}=2.2$ and 25 percent difference for $\alpha_{\rm p}=2.1$, i.e., $\zeta_{\rm prim, no-cutoff}=1.25\zeta_{\rm prim}$ for $\alpha_{\rm p}=2.1$) regardless of whether the exponential terms are included.
Here, $\alpha_{\rm p}=2.1$ is the injection spectrum slope (assumed to be the same for electrons and protons) and $\epsilon^{\rm inj}_{\rm e/p}$ is the energy injected into electrons or protons \citep{werhahn_cosmic_2021a}.\\
\indent
While the value of $(m_{\rm p}/m_{\rm e})^{2-\alpha_{\rm p}}$ is relatively certain, the value of $\zeta_{\rm prim}$ may fluctuate due to different CR acceleration efficiency for CR protons and electrons.  
One-zone model simulations \citep{lacki_physics_2010} reveal that $2$ percent of the kinetic energy of supernova explosion should go to primary CR electrons and $\sim$10-20 percent to CR protons to fit the FIR-radio correlation, which gives $\zeta_{\rm prim}=0.1-0.2$ and corresponds to $K^{\rm inj}_{\rm ep}=0.024-0.048$ for $\alpha_{\rm p}=2.2$ and $K^{\rm inj}_{\rm ep}=0.059-0.12$ for $\alpha_{\rm p}=2.1$. On the other hand, cosmological simulations \citep{ponnada_synchrotron_2024} adopt even smaller values as $\zeta_{\rm prim}=0.02$, which corresponds to $K^{\rm inj}_{\rm ep}=0.0048$ for $\alpha_{\rm p}=2.2$ and $K^{\rm inj}_{\rm ep}=0.012$ for $\alpha_{\rm p}=2.1$. In conclusion, the values of $\zeta_{\rm prim}$ can vary from 0.02 up to 0.2, and $K^{\rm inj}_{\rm ep}$ that we adopt corresponds to $\zeta_{\rm prim}=0.15$ and falls within a reasonable range of $\zeta_{\rm prim}$.\\
\indent
In practice, we normalize the injection spectrum for protons and electrons at their momentum bins that are closest to a kinetic energy of 10~GeV, which is 9.63 and 9.08~GeV for protons and electrons, respectively, due to the limited resolution when modeling the spectrum in momentum space. This small offset leads to a correction factor of $(T_{\rm e, code}/T_{\rm p,code})^{-\alpha_{\rm p}}(1+\alpha_{\rm p}m_{\rm p}c^2/T_{\rm p, code})$, where the second term in brackets comes from a Taylor expansion of the normalized CR momentum with $m_{\rm p}c^{2}/T_{\rm p, code}$ serving as a smallness parameter. Therefore, $K^{\rm inj}_{\rm ep}$ used in the code is $K^{\rm inj}_{\rm ep, code}=K^{\rm inj}_{\rm ep}(9.08/9.63)^{-\alpha_{\rm p}}[1+\alpha_{\rm p}(0.94/9.63)]=$0.13.
\subsection{Radio Emission}
Radio emission comprises of non-thermal synchrotron emission and thermal free-free emission. Here we outline how we model these radiative processes to understand their contributions to the overall radio emission.

\subsubsection{Non-thermal Synchrotron Emission}
We compute the synchrotron emission following \textcolor{blue}{Rybicki \& Lightman (1986)} as 
\begin{equation}
    j_\nu=\frac{\sqrt{3}e^3B_\perp}{m_{\rm e} c^2}\int^{\infty}_0f_{\rm e}(p_{\rm e})F(\nu/\nu_c)\dd p_{\rm e}, 
    \label{eq:syn_j}
\end{equation}
where $e$ is the elementary charge, $B_\perp$ is the magnetic field perpendicular to the line of sight, $f_{\rm e}=\dd N_{\rm e}/(\dd E_{\rm e}\dd V)$ is the CR electron spectral density (i.e., number of particles per unit volume and unit energy), $\nu$ is the observed frequency, and $F(x)\equiv x\int_x^\infty K_{5/3}(\zeta)\dd \zeta$, where $K_{5/3}(\zeta)$ is the modified Bessel function of order 5/3. We adopted an analytical approximation to $F(\nu/\nu_{\rm c})$ provided by \cite{aharonian_angular_2010}. The role of the function $F(\nu/\nu_{\rm c})$ is a synchrotron emission kernel of an individual electron that is centered on the typical synchrotron emission frequency $\nu_{\rm syn}\approx 2\nu_{\rm c}$ for a given energy of a CR particle, where the critical frequency $\nu_{\rm c}$ is defined as
\begin{equation}
    \nu_{c}=\frac{3eB_\perp}{4\pi m_{\rm e} c}\gamma_{\rm e}^2, 
    \label{eq:syn_nu}
\end{equation}
where $\gamma_{\rm e}=\sqrt{p_{\rm e}^2+1}$ is the Lorentz factor corresponding to non-dimensional CR electron energy. This implies that observations of synchrotron emission at a fixed frequency corresponding to lower magnetic field strength probe electrons with higher energy and thus lower CR electron number density.

\subsubsection{Thermal Free-Free Emission}
In addition to synchrotron emission, we also expect thermal bremsstrahlung to contribute to the radio band. We follow \citet{RybickiLightman} and calculate the free-free emissivity from 
\begin{equation}
    j_{\nu, \rm ff} = 6.8\times 10^{-38}
    Z^2\frac{n_{\rm e}n_{\rm i}}{\rm cm^{-6}}\left(\frac{T}{{\rm K}}\right)^{-0.5}e^{-h\nu/(k_{\rm B}T)}{\rm erg s^{-1}cm^{-3} Hz^{-1}}
\end{equation}
where $Z$ is the atomic number, $n_{\rm e}, n_{\rm i}$ is the thermal electron and ion density respectively, and $T$ is the thermal gas temperature. Unlike in \cite{werhahn_cosmic_2021c}, who assume constant gas temperature of $8000$~K, we adopt the thermal gas temperature that is self-consistently calculated in the CRMHD simulation.

\subsubsection{Generating Mock Observation}
In addition to emission, we also consider free-free absorption and synchrotron self-absorption (see Appendix A in \cite{werhahn_cosmic_2021c}) when calculating the final emission maps. We calculate the mock observation with the following procedure: (i) rotate the whole simulation domain to the desired inclination, (ii) calculate the magnetic field perpendicular to line of sight and the resulting emissivity for non-thermal and thermal emission, (iii) divide the simulation cube into thin slices perpendicular to the line of sight, and calculate absorption coefficients for each slice, and finally (iv) calculate the final emission modified by absorption using the radiative transfer equation:
\begin{equation}
    4\pi I_\nu=\int_{\rm s_{0}}^0j_\nu(s)\exp{\left[-\tau_\nu(s,0)\right]} \dd s, 
\end{equation}
where $s$ is the distance to the observer along the line of sight, $s_{0}$ is the distance to the far end of the computational domain, $j_\nu=j_{\rm \nu, syn}+j_{\rm \nu, ff}$ is the total emissivity, $\tau_{\nu}(s,0)=\int_{s}^{0}(\kappa_{\rm\nu, syn}+\kappa_{\rm\nu, ff})\dd s$ is the optical depth, and $\kappa_{\rm syn/ff}$ is the absorption coefficient for synchrotron and free-free emission. We include synchrotron emission and absorption from both primary and secondary CR electrons.\\
\indent
Finally, in order to account for the finite angular resolution of observations and to compare our results to the data for NGC 4217 \citep{stein_chang-es_2020}, we convolve the resulting emission with the \texttt{imsmooth} function in the \texttt{CASA} package \citep{mcmullin_casa_2007} adopting the beam of 13.4~arcsec in diameter (half power beam width), which corresponds to $1.02~{\rm kpc}$ appropriate for this galaxy at a distance of 15.85$~{\rm Mpc}$ \citep{kourkchi_cosmicflows-4_2020}.

\subsection{Calculation of Stokes Parameters}
In order to compare the polarized intensity and the observed magnetic field orientation with the observation, we need to calculate the Stokes $Q$ and $U$ parameters including the Faraday rotation. In this section, we outline the procedure for this calculation, which closely follows that in \cite{waelkens_simulating_2009}.\\
\indent
First, we calculate the polarized emissivity in each cell using the formula in 
\citet{RybickiLightman}, which is similar to Equation~\eqref{eq:syn_j} for the total emissivity except that it contains a different synchrotron kernel:
\begin{equation}
    j_{\nu, \rm pol}=\frac{\sqrt{3}e^3 B_\perp}{m_{\rm e}c^2}\int^\infty_0f_{\rm e}(p_{\rm e})G(\nu/\nu_c)\dd p_{\rm e}, 
\end{equation}
where $G(x)=xK_{2/3}(x)$ is the dimensionless synchrotron kernel function and $K_{2/3}$ is the modified Bessel function of order $2/3$. Second, we calculate the Stokes $Q$ and $U$ emissivities from 
\begin{align}
    j_Q &= j_{\nu, \rm pol}\cos{(2\chi)}, \\
    j_U &= j_{\nu, \rm pol}\sin{(-2\chi)},
\end{align}
where $\chi$ is the polarization angle of each computational cell given by
\begin{equation}
    \chi = {\rm RM}\lambda^2+\chi_0, 
\end{equation}
where $\chi_0$ is the magnetic field direction in each computational cell, $\lambda$ is the wavelength corresponding to the measured frequency $\nu$, and the rotation measure RM is defined as
\begin{equation}
    {\rm RM} = \frac{e^3}{2\pi m_{\rm e}^2c^4}\int_{s_0}^0 n_{\rm e} B_\parallel \dd s,
    \label{eq:rm}
\end{equation}
where $B_\parallel$ is the signed magnetic field parallel to the line of sight and $s_0$ is the distance to the observer. Finally, we integrate the emissivities to obtain Stokes parameters from
\begin{align}
    Q &=\int_{s_0}^{0} j_Q \dd s \\
    U &=\int_{s_0}^{0} j_U \dd s.
\end{align}
The polarized intensity is given by $\sqrt{Q^2+U^2}$ and the observed polarization angle $\phi$, which is the orientation of the magnetic field projected onto the plane of the sky, is defined as
\begin{equation}
    \phi = \frac{1}{2}\arctan{\frac{U}{Q}}.
\end{equation}

\begin{figure*}
    \centering
    \includegraphics[width=\linewidth]{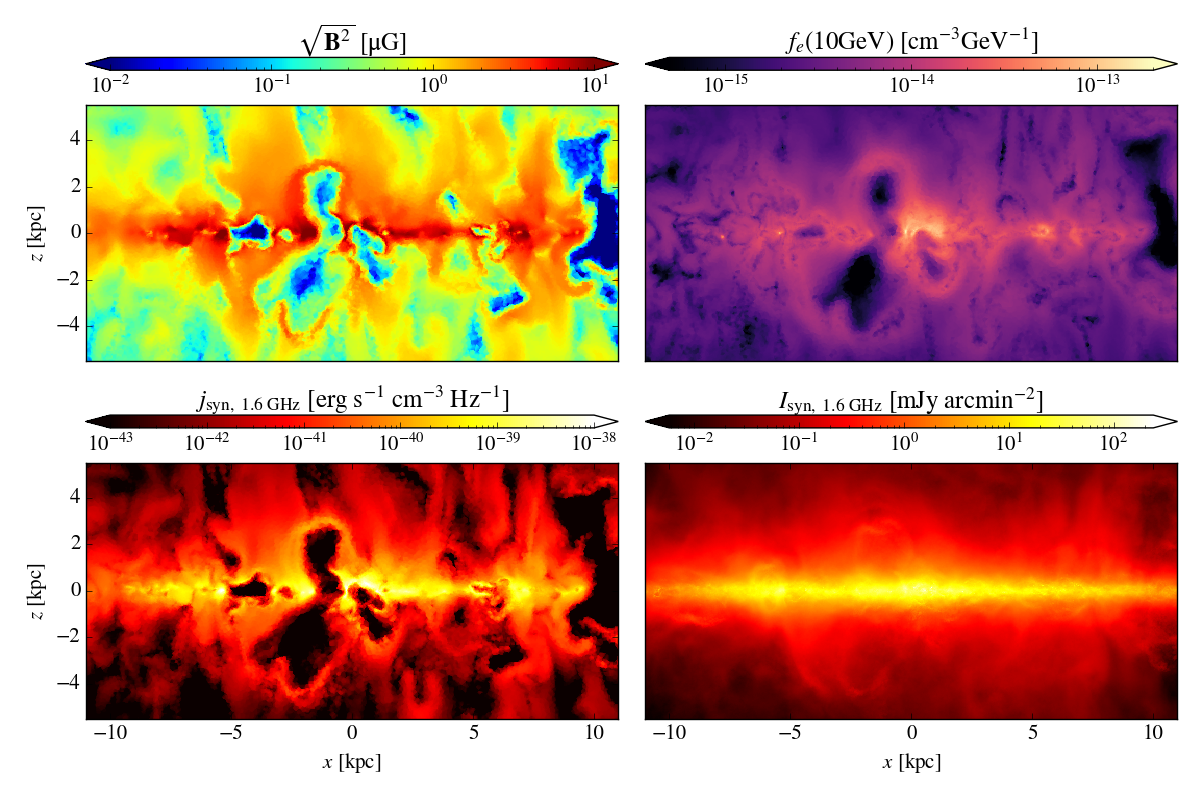}
    \caption{
    Edge-on view of the slice plot (line of sight thickness of 0.5~kpc) of magnetic field magnitude (upper left), CR electron energy density at $10~{\rm GeV}$ (upper right), synchrotron emissivity (lower left) and line-of-sight integrated synchrotron intensity (lower right, projection thickness of 44~kpc). The lower right panel assumes that the galaxy is located at the same distance as NGC~4217 (15.85~Mpc). This plot shows how the magnetic field and the CR electron spectrum contribute to the synchrotron emissivity and the projected intensity, both calculated at 1.6~GHz (L-band) and including contributions from both primary and secondary CR electrons.}
    \label{fig:4panel}
\end{figure*}

\begin{figure*}
    \centering
    \includegraphics[width=\linewidth]{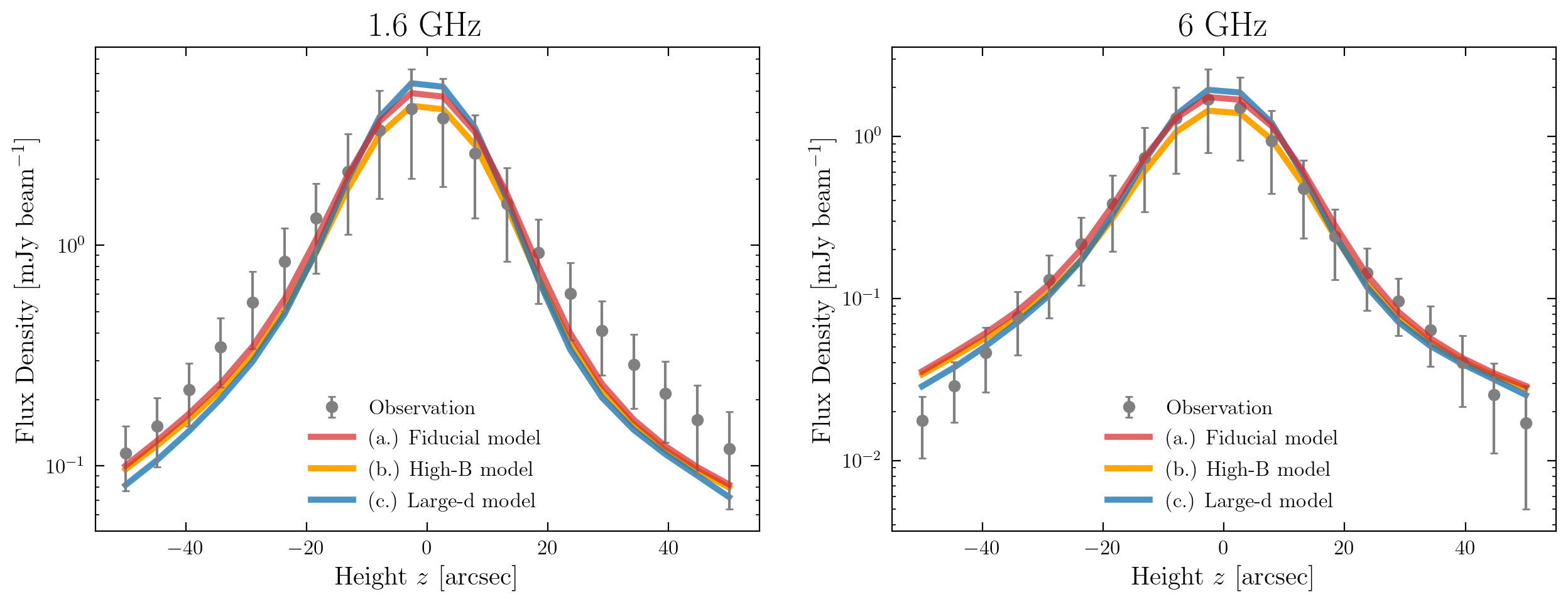}
    \caption{Stacked profile comparison at 1.6~GHz (L band, left panel) and 6~GHz (C band, right panel). Shown are observational data points (grey points with error bars) along with profiles corresponding to 
    (a.) the fiducial model, (b.) high-B model with B field boosted by a factor of 2 in addition to the factor of $1.5^{1/2}$ to account for the difference in SFR between simulation and observations, and (c.) large-d model in which the simulated galaxy is placed at a larger distance ($d$=18.88~Mpc; see legend embedded in the figure). The \textit{magnitude} of the flux density in mock observations that match the data is degenerate with respect to the combination of magnetic field strength, assumed distance, and the normalization of the CR electron spectrum. By contrast, the \textit{shape} of the profile is insensitive to the magnetic field and the assumed distance, and does not depend on the normalization of the CR electron spectrum. We adopt the same beam size area as in \citet{stein_chang-es_2020}, who use the beam size of 203.5~arcsec$^2$.}
    \label{fig:emission_compare}
\end{figure*}

\section{Results} \label{sec:result}

In this section, we compare mock intensity profiles, radio spectra, and morphologies of radio intensity and polarization to the observations of NGC 4217. This galaxy has a slightly higher star formation rate  (SFR; $1.5~{\rm M}_\odot\ {\rm yr}^{-1}$, see \citealt{stein_chang-es_2020}) than the simulated galaxy ($1.0~{\rm M}_\odot\ {\rm yr}^{-1}$, see Fig.~5 in \citealt{thomas_2024arXiv}). Higher SFR in the observation generates (i) more CRs, producing more radio emission, and (ii) more turbulence in the galaxy, which amplifies the magnetic field. Therefore, in the following analysis, we calculate the CR electron spectrum by assuming that the increase in SFR leads to increases in both the CR energy density $e_{\rm CR}$ and magnetic energy density $u_B \propto B^2$ by 1.5, the latter of which increases the magnetic field by $1.5^{1/2}$. We label this as our "Fiducial model" in the following discussion.

\subsection{Radio Intensity Profiles} \label{sec:profile}
We first show in Fig.~\ref{fig:4panel} how the magnetic field and CR electron spectrum contribute to the synchrotron emissivity. 
Overall, the magnetic field strength (upper left panel) has similar morphology to that of the CR electron spectrum at 10 GeV (upper right panel). In particular, a bubble-like feature near $x=-2$~kpc is seen in the distribution of both quantities.
However, the magnetic field exhibits more disk-like morphology compared to the spatial distribution of the CR electron spectrum at 10~GeV.
These features are reflected in the lower-left panel, where we show a slice plot of synchrotron emissivity. The dim bubble-like feature near $x=-2$~kpc 
corresponds to weak magnetic fields and low normalization of the CR electron spectrum, while the bright disk is mainly associated with the strong magnetic field strength. 
Finally, we integrate the synchrotron emissivity over 44~kpc along the line of sight (i.e., over the whole galactic disk) and present the resulting intensity in the lower right panel. The bubble-like feature in the slice plot is covered by synchrotron emission from the outflow at different depths along the line of sight and the intensity distribution is relatively smooth and decreasing as a function of the distance from the galactic midplane.\\
\indent 
In order to quantitatively compare outflow properties to observations, we adopt an approach similar to that in \cite{stein_chang-es_2020} and compute the total intensity profile as a function of height in vertical strips centered on the disk.
Specifically, five strips are set to be perpendicular to the disk with 20 boxes in each vertical strip (see Fig. \ref{fig:strip-define}). The size of the box is set to be the same as in \cite{stein_chang-es_2020}, which has a height of 5~arcsec and width of 35~arcsec, corresponding to 0.4~kpc and 2.7~kpc, respectively. The intensity profile as a function of height $z$ is measured by taking the average intensity within each box. The resulting intensity profiles are shown in Appendix \ref{app:strip} (see Fig.~\ref{fig:strip-Lband} and Fig.~\ref{fig:strip-Cband}). Finally, we construct the overall profile for all strips together and plot the profile.\\
\indent
We show the mean intensity (red line) as a function of height in our fiducial model along with the observational data points (gray points with error bars) in Fig.~\ref{fig:emission_compare}. Both the simulated and observed profiles have a peak at $z=0$, where the galactic disk is located, and decrease by $\sim$1.5 orders of magnitude toward higher latitudes ($\sim$50~arcsec).
While the intensity profile agrees with the observations within the error bars, the agreement is degenerate with respect to various factors--the normalization of the CR electron spectrum, the magnetic field strength, and the assumed distance to the observed galaxy. Below, we discuss the effect of these factors with two different models in addition to our fiducial model (see Table \ref{tab:param} for model parameters).

\begin{table}[]
\centering
\caption{
Model parameters and their values. The magnetic field in all models is increased by a factor of $1.5^{1/2}$ to compensate for the discrepancy between the SFR in the simulation and observations. In the high-B model, an extra factor of 2 is introduced to address the potential underestimation arising from the exclusion of the CGM. In the large-d model, the simulated galaxy is placed at a larger distance (see text for details).}
\begin{tabular}{lllll}
Model    & B field boost & Assumed distance & $K^{\rm inj}_{\rm ep}$ & $\zeta_{\rm prim}$ \\ \hline
Fiducial & $1.5^{1/2}$          & 15.85~Mpc         & 0.092              & 0.15               \\
High-B   & $1.5^{1/2}\times 2$  & 15.85~Mpc         & 0.024              & 0.04               \\
Large-d  & $1.5^{1/2}$          & 18.88~Mpc         & 0.12               & 0.2               
\end{tabular}
\label{tab:param}
\end{table}

First, the uncertainty in the electron-to-proton injection number ratio $K^{\rm inj}_{\rm ep}$ only affects the normalization rather than the shape of the profile. Since increasing $K_{\rm ep}^{\rm inj}$ leads to a higher normalization of the spectrum of primary CR electrons, it only changes the synchrotron emission from \textit{primary} CR electrons. As the \textit{secondary} synchrotron emission and free-free emission are at least three times and three orders of magnitude smaller than the primary synchrotron emission, respectively, the total simulated emission is dominated by primary synchrotron emission. This implies that increasing the value of $K_{\rm ep}^{\rm inj}$ will not change the shape of the total radio emission profile but only increase its normalization.
The prediction of dominant synchrotron emission is consistent with the non-thermal fraction estimated from observations \citep{stein_chang-es_2020}, but the exact synchrotron-to-free free emission ratio depends on detailed modeling of the ISM properties.

Second, the magnetic field in our simulation could be underestimated due to the lack of a realistic CGM. 
In isolated galaxy simulations, the CGM is completely specified by the galactic wind, whereas in cosmological simulations, the CGM originates both from the gas accreted from large distances and the gas expelled by galactic winds over time. 
If there is a pre-existing and structured CGM, localized pockets of velocity shear will develop in the wind resulting in turbulence which would amplify the magnetic field.
For example, \cite{ponnada_synchrotron_2024} also study Milky Way--mass galaxies but in the context of cosmological zoom-in simulations. They obtain a magnetic field magnitude of $\sim$10~$\mu$G at the disk and $\sim$1~$\mu$G at $z=4$~kpc, which is a factor of 2 to 3 larger than our magnetic field magnitude.

We examine the effect of increasing the magnetic field with our "High-B model" (shown as the orange line in Fig.~\ref{fig:emission_compare}). In this model, the magnetic field is boosted by a factor of 2 before post-processing. We find that the $K^{\rm ep}_{\rm inj}$ needed to fit the observation is 0.024, corresponding to $\zeta_{\rm prim}=0.04$, which is a factor of 0.3 smaller than the value in our fiducial model but still within the reasonable range given in the literature \citep{ponnada_synchrotron_2024, lacki_physics_2010}. This implies that a boost in the magnetic field by a factor of 2 leads to an increase in the radio intensity by a factor of $\sim3.4$ and shows that the impact of the underestimation of the magnetic field on radio intensity can be significant. In steady state, the dependence of the radio emissivity on the magnetic and radiation field is \cite[see, e.g., ][]{pfrommer_2008,ruszkowski_cosmic_2023}
\begin{equation}
    j_\nu\propto \frac{B^{\alpha_\nu+1}}{B_{\rm ph}^2+B^2}, 
\end{equation}
where $j_\nu$ is the synchrotron emissivity, $B$ is the magnetic field strength, and $B_{\rm ph}^{2}=8\pi u_{\rm ph}$, where $u_{\rm ph}$ is the energy density in the photon field with contributions from the interstellar radiation field and the CMB.
The observed radio spectral index $\alpha_\nu$ is related to the injected electron spectral index $\alpha_{\rm inj}$ as $\alpha_\nu=\alpha_{\rm inj}/2=1.05$, yielding $j_\nu \propto B^{2.05}$ if $B \ll B_{\rm ph}$. 
Before boosting, $B_{\rm ph}$ is larger than $B$. As we increase the magnetic field, it becomes comparable to $B$ at $z\sim4$~kpc and smaller than $B$ at $z\sim1$~kpc. Therefore, the dependence of $j_\nu$ on $B$ becomes weaker than $j_\nu \propto B^{2.05}$.
This explains why the increase in the radio intensity near the disk midplane is slightly weaker than in the wings of the profile as the field strength is increased.
In practice, we find that the dependence of the radio emission on the magnetic field is super-linear, and hence the magnitude of the radio intensity can be boosted significantly by increasing the magnetic field.

Although the magnitude of the intensity profile depends on the magnetic field, the shape of the profile is relatively insensitive to it. Comparison between the red and orange line in Fig.~\ref{fig:emission_compare} demonstrates that changing the magnetic field strength results in only a slight difference in the shape of the intensity profile. Although the radio intensity is relatively sensitive to the normalization of the magnetic field strength, the match between the emission profile shapes and the observations is a robust feature of our model.

Finally, the assumed distance to the observed galaxy also changes the intensity profile. While our fiducial model adopts the latest measurement from the Cosmicflows-4 collaboration \citep{kourkchi_cosmicflows-4_2020}, the intrinsic scatter of the Tully-Fisher relation could be as high as 20\%. We examine the effect of adopting a larger distance (distance $d=18.88$~Mpc) as measured by the Cosmicflows-3 collaboration \citep{tully_cosmicflows-3_2016} with the blue line in Fig.~\ref{fig:emission_compare}. Although the flux density is independent of distance, 
the physical size of a galaxy characterized by the fixed observed angular size in the sky is larger when the galaxy is placed at a larger distance.
Thus, placing the galaxy at a larger distance effectively results in the steepening of the intensity profile because then a larger fraction of the physical extent of the galaxy,
and hence its dimmer part, fits in the fixed (angular) field of view.
We find that the $K^{\rm ep}_{\rm inj}$ needed to fit the observation is 0.12, corresponding to $\zeta_{\rm prim}=0.20$, which is the maximum value of $\zeta_{\rm prim}$ given in \cite{lacki_physics_2010}. Noticing that the blue line is still consistent with the observation within its error bar, we conclude that the intensity profile is still robust under the 20\% change in the distance.
In conclusion, the shape of the intensity profiles is only weakly sensitive to variations in the magnetic field boost and the assumed distance, making the shape a robust feature of our simulation. However, the normalization of the intensity profile is degenerate with respect to factors such as the magnetic field magnitude, galactic distance, and $\zeta_{\rm prim}$. Thus, a reasonable fit to the intensity profiles (and the radio spectrum) can be achieved by choosing an appropriate combination of these factors. Nonetheless, a detailed model fitting to the data is beyond the scope of this study.

\begin{figure*}
    \centering
    \includegraphics[width=\linewidth]{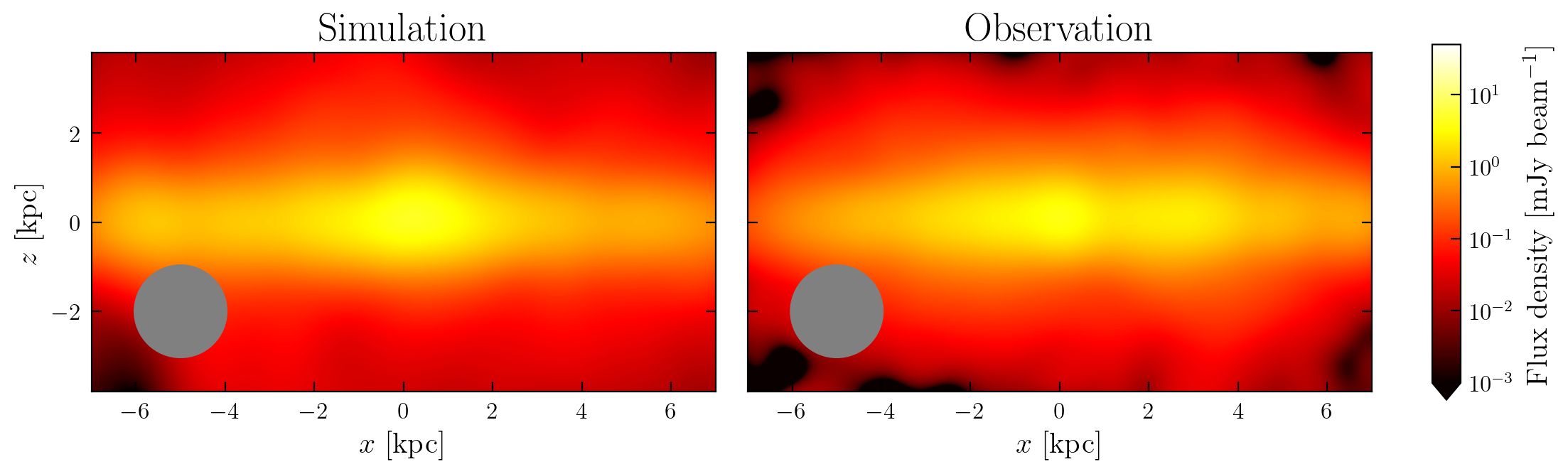}
    \caption{C-band (6~GHz) emission map based on the simulation (left panel) and observation (right panel). The simulated emission map looks similar to the observed image. Both images are smoothed with a beam size of 13.6~arcsec in diameter, whose relative size is shown in the lower-left corner of both panels. The simulated emission map is calculated with our fiducial model.}
    \label{fig:compare_map}
\end{figure*}
\par While the simulated intensity profile with boosted magnetic field is generally consistent with the observational data points within their error bars, the profiles slightly overestimate the intensity near the profile peak at 1.6~GHz and the intensity in the wings at 6~GHz. There could be several reasons for these deviations.
If CR acceleration efficiency is reduced, this would result in a decrease in the CR pressure gradient, thus weakening the galactic outflow and making the intensity profile more concentrated toward the midplane. This might solve the discrepancy at 6 GHz. 
At the same time, this process would also steepen the profiles at 1.6~GHz resulting in a worse match to observations.
However, free-free absorption in the disk can help to reduce the peak at $z=0$~kpc. Although the current optical depth due to free-free absorption is $\lesssim 1$ and does not play an important role in determining the final intensity, increasing the resolution of the simulation at the disk may decrease the gas temperature \citep{hummels_impact_2019, nelson_resolving_2020, esmerian_thermal_2021} and thus lead to an increase of free-free absorption. Since the free-free absorption coefficient $\propto \nu^{-2}$, this effect will be more important at lower frequencies (and at low heights above the disk) and might thus help to cure the small deviation from observations at 1.6~GHz near the peak of the profile while not significantly affecting the emission at 6 GHz.
Therefore, the exact shape of the profiles may depend on various details of the model. 

\subsection{Morphology of Radio Emission} \label{sec: radio emission morph}

\begin{figure*}
    \centering
    \includegraphics[width=\linewidth]{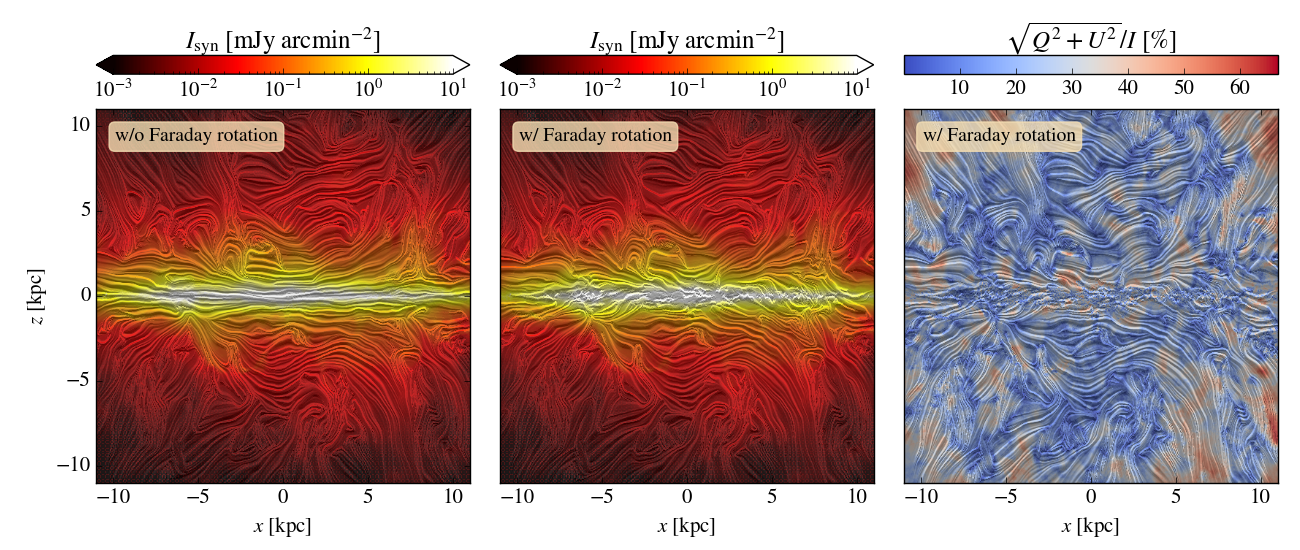}
    \caption{
    Magnetic field direction calculated from Stokes Q and U maps overlaid on Stokes I map (left and middle panel) and magnetic field direction overlaid on the map of polarization degree (right panel). Faraday rotation is included in the middle and right panel when calculating the Stokes Q and U maps. The B field direction shows an X-shape pattern at $|z|>1$~kpc, indicating the existence of an outflow. The polarization degree tends to be higher in regions where magnetic fields are more ordered and vice versa.}
    \label{fig:polarization_3panel}
\end{figure*}

Fig.~\ref{fig:compare_map} shows the total radio emission maps from both observations and simulation at the frequency of 6~GHz with a boosted magnetic field. The right panel shows the radio observation from \cite{stein_chang-es_2020} but rotated to align the disk with the horizontal direction for easier comparison to the simulated mock image shown in the left panel. Both images are smoothed using the same beam size (13.4~arcsec in diameter corresponding to $\sim$1.02~kpc at the distance of interest). Observational noise is added to the simulation by inverse-Fourier transforming the noise spectrum with different random phases but the same amplitude as the observational background noise, which is calculated using the observed intensity distribution at $|z|>4$. After accounting for these observational effects, the simulated image appears blurred to the same extent as the observed one and also exhibits some negative flux regions at the corner of the image.\\
\indent
The simulated emission map displays a wind-like diffuse emission structure that is distributed across the disk midplane, and closely resembles the radio emission pattern observed in real galaxies. This is in contrast to previous works \citep{werhahn_cosmic_2021c, thomas_cosmic-ray-driven_2023}, which exhibit a prominent outflow feature in the central region of the disk at $x=0$. The main difference between our simulation and the above earlier work lies in the initial conditions and the adoption of the \textsc{crisp} ISM model in our work. First, previous studies \citep{werhahn_cosmic_2021c, thomas_cosmic-ray-driven_2023} start with a gravitationally collapsing gas cloud embedded in dark matter halo and simulate the disk formation process without pre-seeded star particles. 
Instead, we post-process the simulations of \citet{thomas_2024arXiv} that start from an idealized setup with star particles seeded randomly following the exponential disk profile. This idealized disk setup allows the simulation to relax in a shorter time and can be run at two orders of magnitude higher mass resolution than the simulation shown by \cite{thomas_cosmic-ray-driven_2023}. Therefore, our simulation represents a later evolutionary stage of the galaxy. Second, compared to earlier studies that rely on the \cite{springel_cosmological_2003} ISM model (but without adopting the phenomenological wind model), our implementation of the much more realistic \textsc{crisp} ISM model enables gas cooling to temperatures as low as several hundred Kelvin, leading to increased star formation activity and preventing the formation of outflows that are too concentrated toward the galactic center. As a result, the edge-on morphology of the radio intensity closely resembles real observations.

\subsection{Morphology of the Polarized Emission}
The presence of the outflow in the simulated galaxy is also confirmed by the X-shaped magnetic field orientation \citep{2020A&A...639A.112K} shown in Fig.~\ref{fig:polarization_3panel}. 
Due to magnetic flux freezing, the field lines extend from the disk into the halo as the wind develops and the gas flows away from the disk.
The magnetic field direction without and with Faraday rotation considered are shown in the left and middle panels, respectively. 
As shown in Equation~\eqref{eq:rm}, the rotation measure is proportional to the thermal electron density and the magnetic field, both of which have larger values in the disk midplane (see top panel of Fig.~\ref{fig:2panel_rm}; the polarization angles can be Faraday rotated in the disk by up to 15~degrees at 6~GHz; see bottom panel of Fig.~\ref{fig:2panel_rm}). Therefore, the rotation measure is highest within the disk and decreases toward the halo. As a result, comparing the left and middle panels in Fig.~\ref{fig:polarization_3panel}, the halo regions ($|z|>1$~kpc), which are relatively unaffected by Faraday rotation, clearly show the X-shape pattern of the magnetic field indicating the existence of an outflow. In the left panel of Fig.~\ref{fig:polarization_3panel}, where the Faraday rotation is not considered, the magnetic field direction near the disk is horizontal due to the projection of the foreground and background field (see Fig.~\ref{fig:2panel_zoomin}).

In the right panel of Fig.~\ref{fig:polarization_3panel}, we present the field direction overlaid on top of the polarization degree with Faraday rotation included. The polarization degree is smaller when the field is more tangled, but it can reach the typical maximum synchrotron polarization degree ($\sim 70-75$ percent for electron spectral index in between 2.1 and 3.1) when the field is more ordered, such as in the X-shape region and just above the disk.\\
\begin{figure*}
    \centering
    \includegraphics[width=\linewidth]{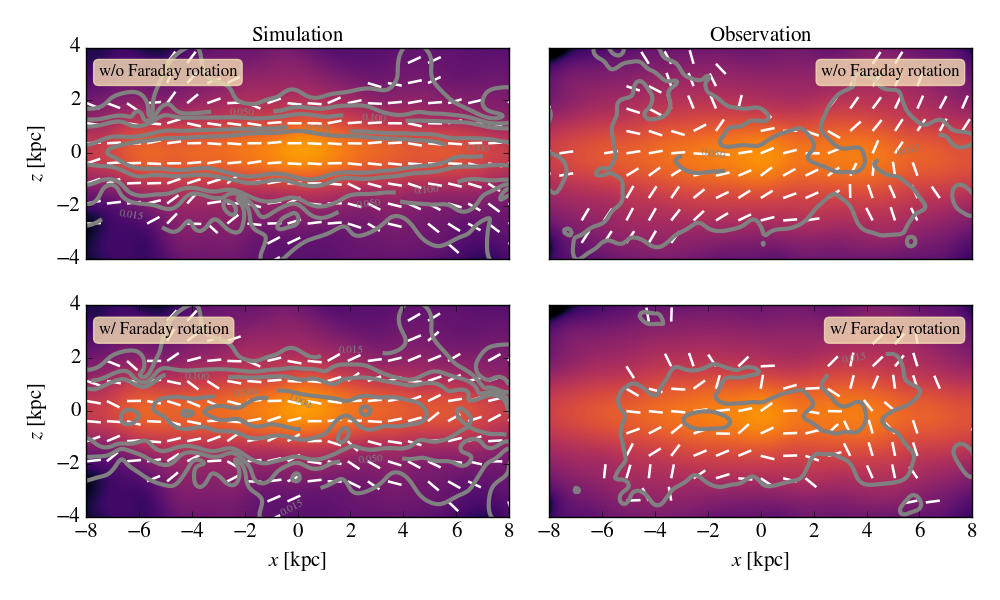}
    \caption{
    Comparison of the X-shaped magnetic field morphology and the polarized intensity between simulation and observation. The contours show polarized intensity at 0.015, 0.050, 0.1, 0.5, and 1 mJy with beam depolarization considered. The white-dashed lines show the magnetic field direction, plotted on top of the total emission map in 6 GHz. The left panel depicts the simulation results, while the right panel shows the real observation (same data as in Figure 11 in \citealt{stein_chang-es_2020}). 
    The upper panel corresponds to the field direction without the influence of Faraday rotation, whereas the lower panel shows the field direction with Faraday rotation included. The magnetic field direction exhibits a characteristic X-shaped pattern indicative of an outflow. In lower panels, the field orientation within the disk region is Faraday-rotated, while it continues to display an X-shaped pattern at $|z|>1$~kpc.
    \label{fig:polarization_4panel}}
\end{figure*}
\indent
Fig.~\ref{fig:polarization_4panel} shows mock observations (left column) and observations (right column). The background shows \textit{total} radio emission convolved to a beam size of 7.7~arcsec ($\sim0.59$~kpc), while the contours show the \textit{polarized} intensity at 0.015, 0.05, 0.1, 0.5, and 1~mJy~beam$^{-1}$ when applicable since the maximum value of the four panels is different. Magnetic field orientations in regions where the polarized intensity is larger than 0.015~mJy~beam$^{-1}$ are shown as white dashed lines. Beam-depolarization is included by smoothing the Stokes Q and U maps with a 7.7~arcsec beam in diameter before calculating the polarized intensity and magnetic field orientation. \\
\indent
The upper (lower) panels show results without (with) Faraday rotation. In the mock observation, observational noise generated by a Fourier transform approach is added to the total intensity shown in the background (in the same fashion as in Section \ref{sec: radio emission morph}). The upper-left panel shows a significantly larger (maximum contour being 1~mJy/beam) polarized intensity than the upper-right panel (maximum contour being 0.05~mJy~beam$^{-1}$). This is in part because the observational image in the upper-right panel is obtained by Faraday de-polarizing the lower-right panel, while the simulation data can be used to directly calculate the polarized intensity without Faraday rotation and, thus, without any information loss. When Faraday rotation is considered (lower left panel), the simulated polarized intensity is still larger (maximum contour being 0.1~mJy~beam$^{-1}$) than the observed polarized intensity because we do not model observational noise for polarized intensity. \\
\indent
In Fig.~\ref{fig:polarization_4panel}, the field orientations in the simulation exhibit a pattern consistent with the observational data. Both the simulation and observation reveal that near the disk the field directions are parallel to the midplane when the Faraday rotation is not considered (upper panel), and become disordered when Faraday rotation is included (lower panel). 
At higher latitudes, the B-field directions are not parallel, but instead slightly tilted or even perpendicular to the midplane, suggesting the presence of an outflow. This X-shaped feature persists even when Faraday rotation is included, because it primarily affects the emission from the vicinity of the midplane. 
Fig.~\ref{fig:polarization_3panel} also shows that the field directions can still be perpendicular to the midplane at even higher latitudes (than those shown in Fig.~\ref{fig:polarization_4panel}), where the polarized intensity is too faint to be detected.

\subsection{Radio Spectrum}
\begin{figure}
    \centering
    \includegraphics[width=\columnwidth]{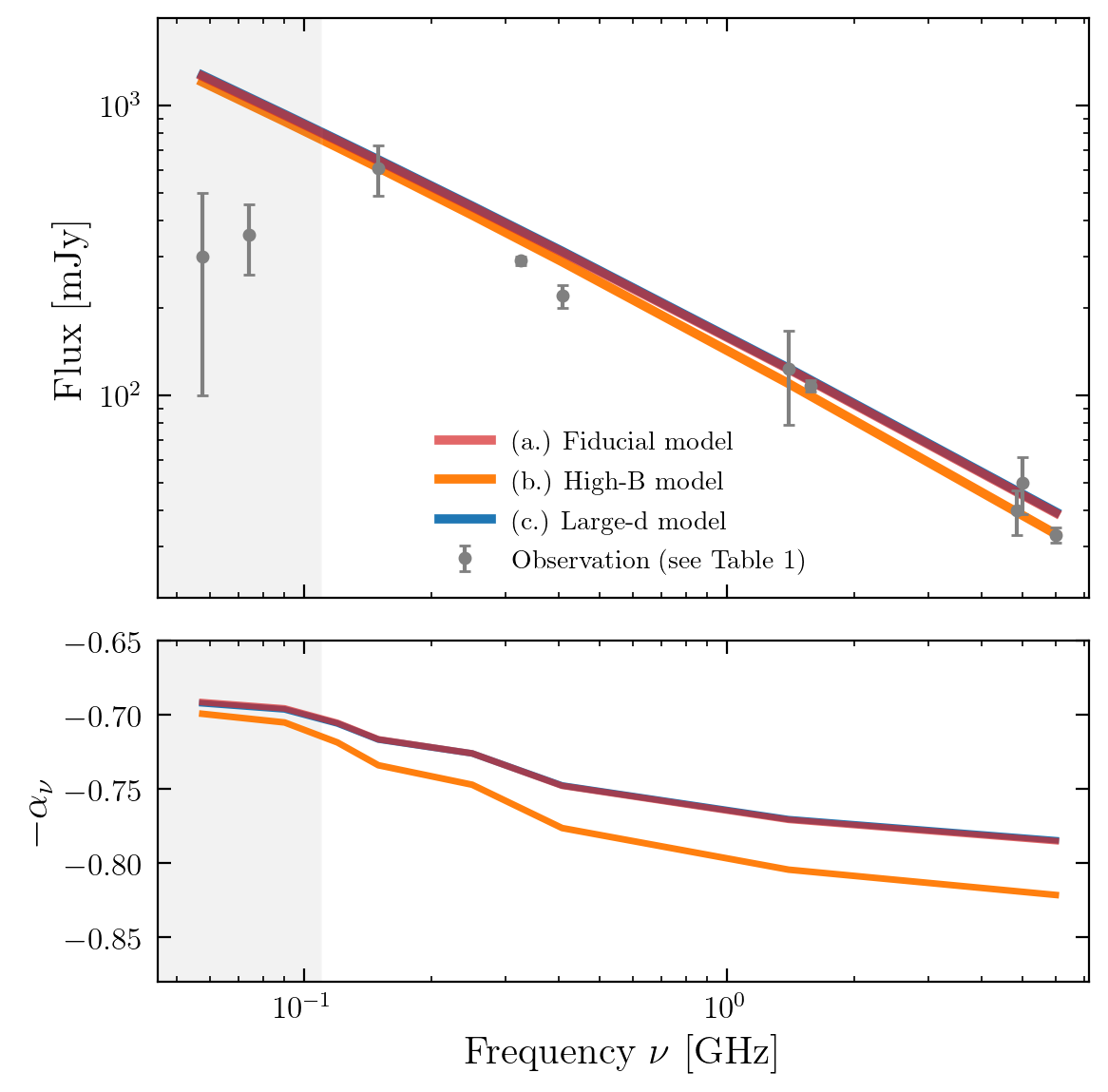}
    \caption{
    \textit{Top panel}: Observed (gray points with error bars) and simulated (solid lines) radio spectrum of NGC 4217. All of the simulated spectra with different parameter configurations are consistent with the observations for $\nu>0.1~ {\rm GHz}$. 
    Both the observational and simulation data for $\nu<0.1$~GHz (shaded region) are less reliable due to the low resolution of the observations at low frequencies and finite numerical resolution when modeling free-free absorption. Observational data and the associated references are given in Table \ref{tab:spectra}.
    \textit{Bottom panel}: Spectral indices of the simulated spectra. Changing distance does not affect the shape of the spectrum while increasing the magnetic field slightly steepens the spectrum. A spectral index larger than $\alpha_\nu=0.7$ indicates that IC and synchrotron cooling processes are important. 
    }
    \label{fig:spectrum}
\end{figure}

We compare the simulated and observed spectra of NGC~4217 in the upper panel of Fig.~\ref{fig:spectrum} and show the spectral indices $\alpha_\nu=-\frac{\dd \log F_\nu}{\dd \log\nu}$ in the lower panel, where $F_\nu$ is the frequency-dependent flux and $\nu$ is the photon frequency. The references for the observational data are listed in Table~\ref{tab:spectra}. 
In order to facilitate a meaningful comparison of the simulations and observations, we include data points from various sources that provide measurements of the \textit{total} flux (i.e., solid angle-integrated) of NGC 4217 and exclude those that only record a \textit{peak} flux.

The spectra predicted for all of our three cases (fiducial, high-B, and large-d) are consistent with the observations for frequencies $\nu>0.1~{\rm GHz}$. A comparison of these simulation-based spectra shows that spectral slope does not depend on the adopted galactic distance and is only slightly steepened in the high-B case due to the increased synchrotron cooling.

Spectral indices in all cases vary from 0.7 to 0.8, indicating the importance of synchrotron and IC cooling for the following reasons. Since we adopt an energy-dependent diffusion coefficient $D(E)=D_0(E/E_0)^{0.3}$, CR losses due to diffusion can steepen the spectrum by 0.3 only, resulting in a CR electron spectrum index $\alpha_{\rm e}=\alpha_{\rm inj}+0.3=2.4$ and the radio spectral index $\alpha_\nu=(\alpha_{\rm e}-1)/2=0.7$. By contrast, IC and synchrotron losses can steepen the injection spectra by unity in steady state \citep{winner_evolution_2019}, which leads to a CR electron spectrum index of $\alpha_{\rm e}=3.1$ and a radio spectral index of 1.05. Therefore, spectral indices larger than 0.7 imply that synchrotron and IC cooling are important.

The consistency between the observations and simulation is non-trivial in terms of the normalization and the slope. On one hand, as discussed in section \ref{sec:profile}, variations in the SFR, magnetic field, and the normalization of the CRe electron spectrum can change the total flux of the simulated galaxy. On the other hand, the slope of the CRe spectrum could be very different with or without the steady-state assumption (see Fig.~15 in \citealt{ruszkowski_cosmic_2023}) resulting in a very different radio synchrotron emission spectrum. 
Therefore, the consistency of the simulated and observed slopes is a testament to the fact that cooling losses play an important role in shaping the spectrum.\\
\indent
Both the observed and simulated spectrum for frequencies $\nu<0.1$~GHz are not reliable. On the observation side, the low frequency (and hence low resolution) measurement from \cite{marvil_integrated_2015} at 74~MHz is determined by measuring the peak flux and multiplying it by the ratio of the solid-angle-integrated flux to peak flux based on a measurement at a higher frequency, which has better resolution and for which the target is resolved. However, the integrated-to-peak flux ratio is not constant, but increases with decreasing frequency. For example, the ratio of solid-angle-integrated flux to the disk flux for NGC 4217 increases from 2.35 at 6~GHz to 5.4 at 0.15~GHz as measured in \cite{stein_chang-es_2020}. Therefore, using the ratio of integrated to peak flux at a higher frequency to infer low-frequency measurement will lead to an underestimate of the real total flux since the actual ratio at low frequency is higher than it is at higher frequency. On the simulation side, while the free-free absorption could significantly attenuate the low-frequency spectrum, the modeling of free-free opacity depends sensitively on the simulation resolution and the detailed ISM modeling. For example, increasing resolution at the disk region may lead to increased cooling and hence more absorption \citep{hummels_impact_2019, nelson_resolving_2020,esmerian_thermal_2021}. In conclusion, our simulation shows consistency with observations for $\nu>0.1$~GHz, while the spectrum at low frequency ($\nu<0.1$~GHz) remains uncertain. 

\begin{table}
\centering
\caption{Total flux of NGC 4217}
\label{tab:spectra}
\begin{tabular}{cll}
Freq. (GHz)  & Flux (Jy)   & Reference              \\ \hline
0.0575  &   0.3$\pm$0.2   & \cite{israel_low-frequency_1990}   \\
0.074   & 0.357$\pm$0.097 & \cite{marvil_integrated_2015}     \\
0.15    & 0.608$\pm$0.121 & \cite{shimwell_lofar_2017}        \\
0.325   & 0.291$\pm$0.009 & \cite{marvil_integrated_2015}            \\
0.408   & 0.220$\pm$0.0205 & \cite{ficarra_new_1985}    \\
1.4     & 0.123$\pm$0.0044 & \cite{condon_nrao_1998}  \\
1.58    & 0.108$\pm$0.005  & \cite{stein_chang-es_2020}             \\
4.85    & 0.04$\pm$0.007   & \cite{gregory_87gb_1991} \\
5       & 0.05$\pm$0.011   & \cite{sramek_5-ghz_1975} \\
6       & 0.033$\pm$0.002  & \cite{stein_chang-es_2020}     
\end{tabular}
\end{table}

\section{Conclusions} \label{sec:conclusion}
In this work, we model the radio emission from an edge-on galaxy using a two-moment CR magnetohydrodynamical simulation \citep{thomas_2024arXiv} and compare the results to the edge-on galaxy NGC~4217 \citep{stein_chang-es_2020}. We compute the spectra of CR protons, primary electrons, and secondary electrons adopting a cell-based steady-state approach \citep{werhahn_cosmic_2021a}. We then calculate thermal and non-thermal radio emission and absorption processes and the Stokes parameters including Faraday rotation and beam depolarization.\\
\indent
We compare observations of the latitude- and longitude-dependent intensity profiles obtained following a procedure similar to that adopted by \cite{stein_chang-es_2020}. 
The intensity profiles at both 1.6 and 6~GHz agree with the observations within the error bars when accounting for the difference in the SFR of the simulated and observed galaxies. This agreement is achieved by simultaneously increasing the CR energy density by 1.5 times and adjusting the magnetic field strength by a factor of $1.5^{1/2}$ to compensate for the underestimation of CR production and turbulence.
We examine the dependence of our results on different factors such as the magnetic field, assumed distance of the observed galaxy, and the normalization of the CR electron spectrum. 
We find that although the magnitude of the intensity profile is degenerate with respect to these model elements, the shape of the profile is only very weakly dependent on these factors and it is thus a robust prediction of our model.\\
\indent
Our simulated galaxy exhibits a morphology similar to that of NGC~4217 in terms of radio emission and magnetic field orientation. The emission map of our simulated galaxy shows a spatially-extended wind-like structure with emission evenly distributed across the mid-plane, which better resembles galaxies that have no centrally biased star formation rate density. Such centrally peaked star formation events can be realized in starburst systems or systems with a strong stellar bar feature.

This extended wind-launching disk naturally arises in (i) our sophisticated ISM \textsc{crisp} model that allows the gas to cool to lower temperatures and to generate more stellar winds and supernovae to prevent star formation at the center of unbarred galaxies, and (ii)  because we effectively simulate later evolutionary stages of the galaxy with higher mass resolution. Our simulations also reveal X-shaped magnetic field patterns at heights $\sim$1~kpc above the disk midplane irrespective of the presence of Faraday rotation. This feature is also present in the observations and indicates the existence of an outflow in both the simulation and observation.\\
\indent
Finally, the simulated spectrum of our galaxy matches that of NGC~4217 for $\nu>0.1$~GHz, while the observed spectrum at lower frequencies remains uncertain. The shape of the spectrum is also essentially insensitive to the magnetic field strength. 

Our findings demonstrate the effectiveness of the two-moment cosmic ray transport model in generating realistic outflows on galactic scales and provide a deeper understanding of galactic wind dynamics and its influence on the evolution of galaxies.

\section*{Acknowledgements}
HHC and MR thank Yelena Stein, Volker Heesen, Jiangtao Li, and I-Hsuan Kuo for their valuable suggestions on observational data processing. 
MR acknowledges support from the National Aeronautics and Space Administration grant ATP 80NSSC23K0014 and the National Science Foundation Collaborative Research Grant NSF AST-2009227.
TT and CP acknowledge support by the European Research Council under ERC-AdG grant PICOGAL-101019746. This work was supported by the North-German Supercomputing Alliance (HLRN) under project bbp00070.
%
%MR 
This research was supported in part by grant NSF PHY-2309135 to the Kavli Institute for Theoretical Physics (KITP).
%
%MR, TT, MW, and CP 
This work was performed in part at the Aspen Center for Physics, which is supported by National Science Foundation grant PHY-2210452, and by grants from the Simons Foundation (1161654, Troyer), and Alfred P Sloan Foundation (G-2024-22395).
MR acknowledges the support from the Leinweber Center for Theoretical Physics which sponsored the organization of the 7th ICM Theory and Computation Workshop during which HHC, MR, and CP were able to finalize key aspects of the project. 
MR thanks Volker Springel and the Max Planck Institute for Astrophysics (MPA) in Garching for their hospitality during his sabbatical stay at MPA. MR acknowledges Forschungsstipendium from MPA. 

\section*{Data Availability}
The data underlying this article will be shared on reasonable request to the corresponding author.

\appendix
\restartappendixnumbering
\section{Verifying the steady-state assumption} \label{app:steady-state}
\begin{figure*}
    \centering
    \includegraphics[width=\linewidth]{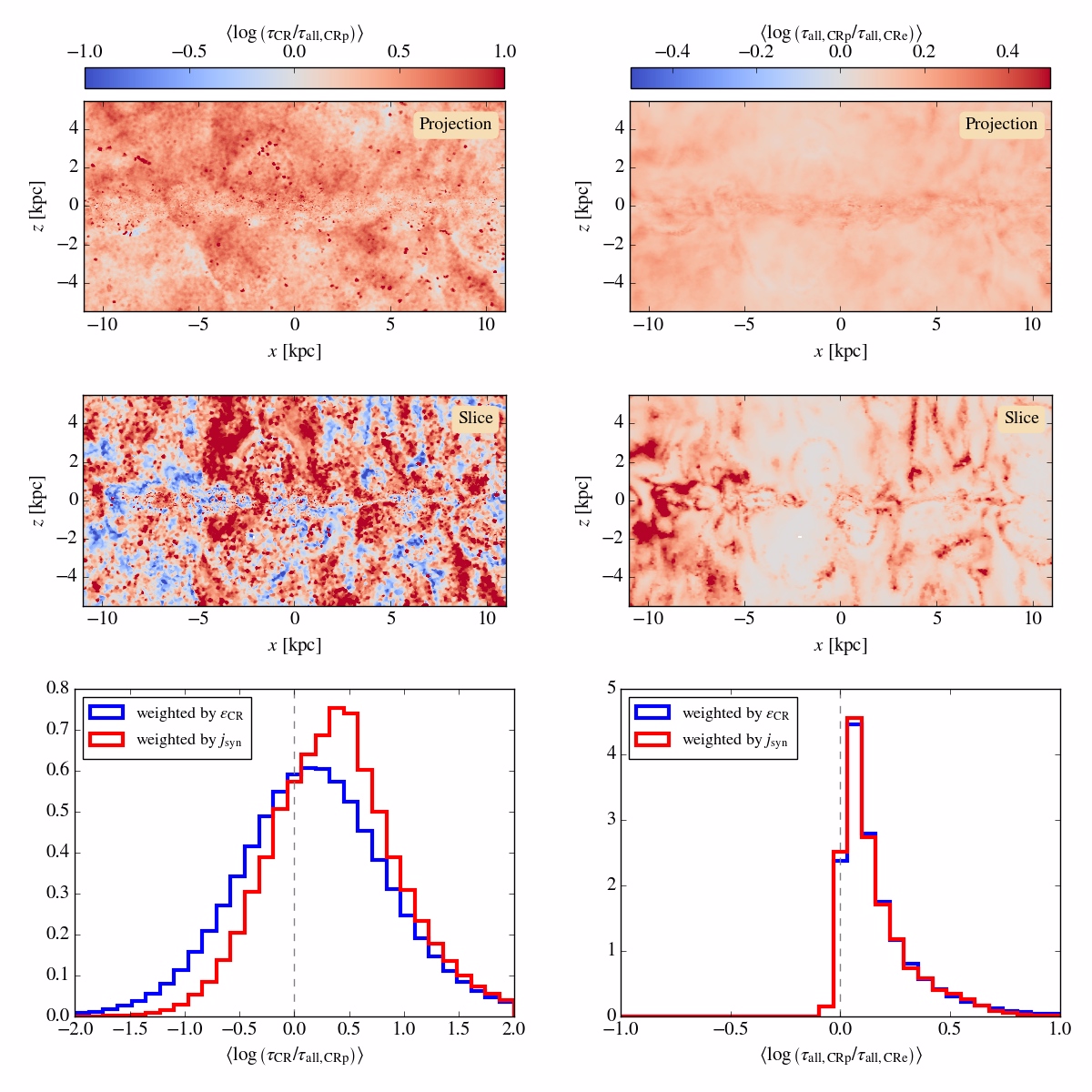}
    \caption{
    {\it Left column}: Adequacy of the steady-state assumption. Upper left panel: projection of the timescale ratio within the whole box ($y=-10$~kpc to $y=10$~kpc; upper panel) and a thin slice perpendicular to the line of sight anchored at $y=0$ (middle panel) both weighted by total radio emissivity; Lower left panel: histogram of weighted timescale ratios.
    The projection plots show that the steady-state assumption is adequate for heights from $z=0$ to $z\pm 5$~kpc (Fig. \ref{fig:emission_compare}), while the histogram shows that the timescale ratio is greater than 1 in cells that contribute most to the radio spectrum (Fig. \ref{fig:spectrum}).
    {\it Right column}: Adequacy of the steady-state assumption for electrons. Upper-right panel: projection of the cooling timescale ratio between protons and electrons within the whole box ($y=-10$~kpc to $y=10$~kpc; upper panel) and a thin slice perpendicular to the line of sight anchored at $y=0$ (middle panel) both weighted by total radio emissivity; Lower-right panel: histogram of weighted cooling timescale ratios.
    The projection plot shows that the steady-state assumption is adequate for CR electrons for our intensity profile and the radio spectrum, assuming that the energy density of electrons is always a fraction of that of protons.
    }
    \label{fig:timescale_combine}
\end{figure*}

The steady state is defined by the requirement that
\begin{equation}
   \dot{e}_{\rm all} + \dot{e}_{\rm inj} \equiv \dot{e}_{\rm CR} = 0,
\end{equation}
where $\dot{e}_{\rm all}$ is the rate of CR energy change that includes contributions from various cooling processes as well as the advection of CRs in and out of the computational cells. That is, $e_{\rm CR}/|\dot{e}_{\rm all}|\equiv \tau_{\rm all}$ includes all CR energy density changes other than an imposed CR injection rate $\dot{e}_{\rm inj}$ that is necessary to keep the CR distribution in a steady state (as assumed in the approach used to post-process the simulations to generate steady-state radiation spectra). Thus, the steady-state solution is exactly correct when $e_{\rm CR}/|\dot{e}_{\rm CR}|$ is infinite. 
In other words, the system is in a near steady state when $|\dot{e}_{\rm all}|\sim |\dot{e}_{\rm inj}|$ or
$|\dot{e}_{\rm all}+\dot{e}_{\rm inj}|\ll(|\dot{e}_{\rm all}|+|\dot{e}_{\rm inj}|)/2$.
In that case, $e_{\rm CR}/|\dot{e}_{\rm all}| = \tau_{\rm all} \ll e_{\rm CR}/|\dot{e}_{\rm CR}| \equiv \tau_{\rm CR}$, and so the system is close to a steady state when $\tau_{\rm all}\ll \tau_{\rm CR}$ in which case the application of the steady-state post-processing technique leads to near exact results.\\
\indent
The CR energy change timescale $\tau_{\rm CR}$ is calculated by comparing the CR energy in each computational cell of two adjacent snapshots separated in time by $\dd t\sim 0.1$~Myr. The energy loss timescale includes the escape timescale $\tau_{\rm esc}$ and the cooling timescale $\tau_{\rm cool}$ and is given by
\begin{equation}
    \frac{1}{\tau_{\rm all}}=\frac{1}{\tau_{\rm esc}}+\frac{1}{\tau_{\rm cool}}, 
\end{equation}
where the escape timescale is a combination of diffusion and advection timescale $\tau_{\rm esc}^{-1}=\tau_{\rm diff}^{-1}+\tau_{\rm adv}^{-1}$ (see Section \ref{sec:model spectra} for $\tau_{\rm diff}$ and $\tau_{\rm adv}$) and the cooling timescale is given by the total energy divided by the proton energy loss rate $\tau_{\rm cool} = e_{\rm CR}/|b_{\rm coul}+b_{\rm pion}|$, where $b_{\rm coul}$ and $b_{\rm pion}$ are the energy loss rates due to Coulomb and hadronic interactions, respectively. We calculate the escape and cooling timescales at CR energy of 10~GeV. In order to verify the cell-based steady-state assumption, we show in the left panel of Fig.~\ref{fig:timescale_combine} the ratio of timescales computed as $\left<\log\left(\tau_{\rm CR}/\tau_{\rm all}\right)\right>$, where the average is intensity-weighted.\\
\indent
We motivate this choice as follows. While $\log(\left<\tau_{\rm all}/\tau_{\rm CR}\right>^{-1})$ is biased toward small ratios and $\log(\left<\tau_{\rm CR}/\tau_{\rm all}\right>)$ is biased toward large ratios, 
 $\left<\log\left(\tau_{\rm CR}/\tau_{\rm all}\right)\right>$ provides a typical in-between value of the timescale ratio and is independent of how we define the timescale ratios (i.e., $\left<\log\left(\tau_{\rm CR}/\tau_{\rm all}\right)\right> = -\left<\log\left(\tau_{\rm all}/\tau_{\rm CR}\right)\right>$). In the absence of intensity weighting, the above logarithms of averages would translate into the logarithm of the harmonic $\left<\tau_{\rm CR}/\tau_{\rm all}\right>_{\rm H}$, geometric $\left<\tau_{\rm CR}/\tau_{\rm all}\right>_{\rm G}$, and arithmetic average $\left<\tau_{\rm CR}/\tau_{\rm all}\right>_{\rm A}$, respectively, which are strictly ordered as $\left<\tau_{\rm CR}/\tau_{\rm all}\right>_{\rm H}\leq\left<\tau_{\rm CR}/\tau_{\rm all}\right>_{\rm G}\leq\left<\tau_{\rm CR}/\tau_{\rm all}\right>_{\rm A}$. In practice, we retain intensity weighting to ensure that the computational cells that contribute most to intensity can properly shape the timescale ratio.\\

\indent
We evaluate timescale ratios in each computational cell and plot (i) a projected map weighted by total radio emissivity in a box (20~kpc on a side; upper panel) and a slice (0.5~kpc thick and perpendicular to the line of sight; middle panel) and (ii) a histogram weighted by CR energy density and total radio emissivity. Although there are some clumpy regions in the slice plot where the $\tau_{\rm CR}/\tau_{\rm all}$ is smaller than 1 (shown in blue), the radio emission-weighted emission average projection shows that the ratio is around $10^{0.5}\sim 3$, meaning that the steady-state assumption is approximately valid in the region where we calculate the intensity profile as a function of height (Fig. \ref{fig:emission_compare}). The histogram, showing the ratio distribution of the most luminous computational cells, verifies the adequacy of the steady-state assumption for the purpose of calculating the radio spectrum (Fig.~\ref{fig:spectrum}).\\
\indent
In order to check the adequacy of the steady-state assumption for electrons, we show in the right panel of Fig.~\ref{fig:timescale_combine} the distribution of the ratio between the energy loss timescale for CR protons, $\tau_{\rm all, CRp}$, and electrons, $\tau_{\rm all, CRe}$, with the same procedure as in the left panel of Fig.~\ref{fig:timescale_combine}. Both the maps and histogram show that the $\tau_{\rm all, CRe}$ is shorter than $\tau_{\rm all, CRp}$, which is generally true for relativistic protons \citep{ruszkowski_cosmic_2023}. In our simulation, the cooling timescale for CRe and CRp are dominated by advective, diffusive, and streaming timescales, which is identical for CRe and CRp. Therefore, $\tau_{\rm all,CRp}\approx\tau_{\rm all,CRe}$ corresponds to regions where advective, diffusive or streaming dominate while the tail of the distribution ($\tau_{\rm all,CRp}>\tau_{\rm all,CRe}$) implies that the radiative leptonic loss rates dominate over losses associated with spatial transport.\\
\indent
Assuming that the energy density of CR electrons is always some fraction of CR protons, $\tau_{\rm CR}$ is then identical for protons and electrons. Therefore, we obtain $\tau_{\rm all, CRe}<\tau_{\rm all, CRp}<\tau_{\rm CR}$ and verify that the steady-state assumption is also fulfilled for CR electrons. All ratios calculated here adopt the original magnetic field strength in the simulation and thus correspond to the lower limit of $\tau_{\rm CR}/\tau_{\rm all, CRe}$ because increasing the magnetic field strength will also increase synchrotron losses and decrease the cooling timescale.

\section{Rotation measure}\label{app:rotation measure}
\begin{figure}
    \centering
    \includegraphics[width=\linewidth]{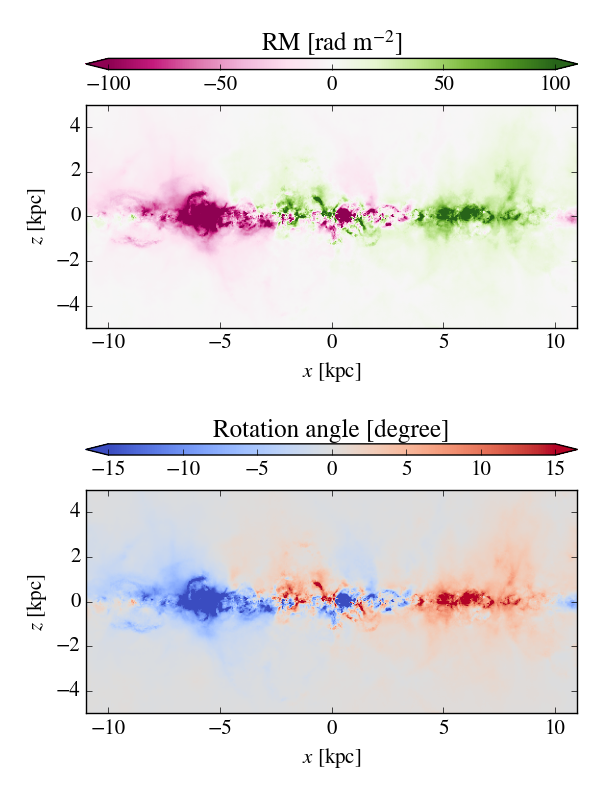}
    \caption{Rotation measure of the simulated galaxy. The top panel shows the Faraday rotation angle per wavelength square, and the bottom panel shows the total rotation angle in degrees at 6 GHz that can be used to compare with Fig.~\ref{fig:polarization_4panel}.}
    \label{fig:2panel_rm}
\end{figure}

We show the rotation measure (RM) of the simulated galaxy in Fig.~\ref{fig:2panel_rm}. The upper panel shows the RM value in units of radian per $\rm m^2$ with RM values near the disk midplane reaching values of $\sim \pm 200\ {\rm rad\ m^{-2}}$. The lower panel shows the total rotation angle of the magnetic field along the integrated line of sight at a frequency of 6 GHz to enable easy comparison to Fig.~\ref{fig:polarization_4panel}, which shows that the field direction is mainly rotated near the disk midplane, where the polarization angles can be rotated at most by $\sim30$~ degrees. Both panels show a transition of the RM values from positive to negative at the galactic center, which is also seen in the RM of the Milky Way \citep{reissl_reproduction_2023}. This transition corresponds to the reversal of the magnetic field direction and hence the toroidal magnetic field structure feature in our simulation. 
\restartappendixnumbering

\section{Origin of the horizontal magnetic field within the disk} \label{app: x-shape field}
\begin{figure}
    \centering
    \includegraphics[width=\linewidth]{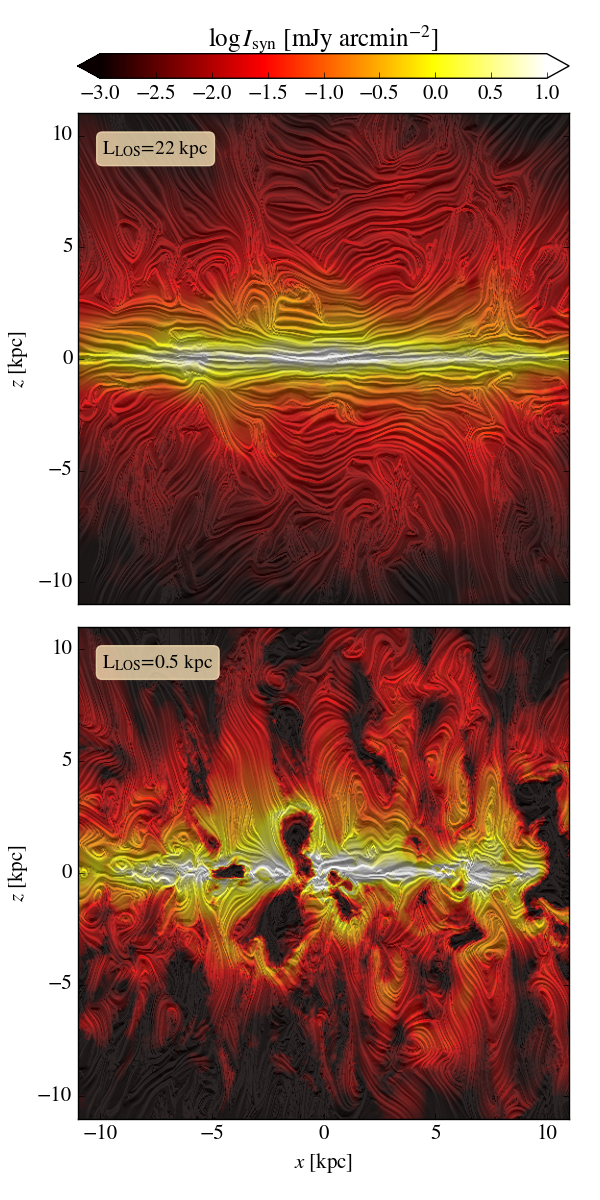}
    \caption{Magnetic field direction calculated from Stokes Q and U maps overlaid on Stokes I map with different integration depth along the line of sight $L_{\rm LOS}$ (Top: 22~kpc, same as left panel of Fig.~\ref{fig:polarization_3panel}; Bottom: 0.5~kpc). The Faraday rotation effect is not included. The field direction in the disk in the lower panel is more structured than in the upper panel, indicating that the horizontal field direction within the disk is due to foreground and background projection.}
    \label{fig:2panel_zoomin}
\end{figure}

In Fig.~\ref{fig:2panel_zoomin}, we show the magnetic field direction on top of the intensity map with different depths of integration. The upper panel is the same as the left panel in Fig.~\ref{fig:polarization_3panel} 
in order to compare with the lower panel, which adopts a shorter integration length along the line of sight. The increased disorder in the field direction observed in the lower panel suggests that the horizontal magnetic field within the disk, depicted in the upper panel, results from the influence of the foreground and background gas.\\

\restartappendixnumbering

\section{Intensity profiles for individual data strips}\label{app:strip}
We show the raw data of the average intensity along five different strips in our simulation to demonstrate the large variation in the direction parallel to the disk. For example, in Fig.~\ref{fig:strip-Lband}, the intensity at $x=0$ increases from around 2 mJy/beam to 10 mJy/beam between strip I and strip III; Fig.~\ref{fig:strip-Cband} also shows about half an order of magnitude variation between strips. 

\newpage
\restartappendixnumbering
\begin{figure}
    \centering
    \includegraphics[width=\linewidth]{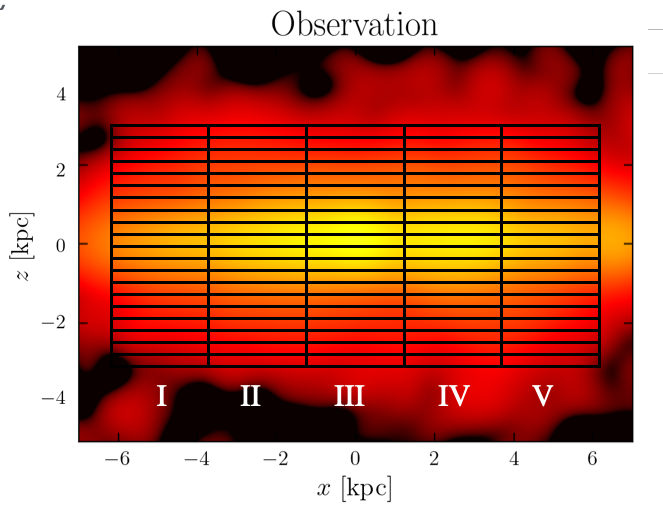}
    \caption{Simulated intensity with five strips defined to fit the intensity profile.}
    \label{fig:strip-define}
\end{figure}
\begin{figure*}
    \centering
    \includegraphics[width=\linewidth]{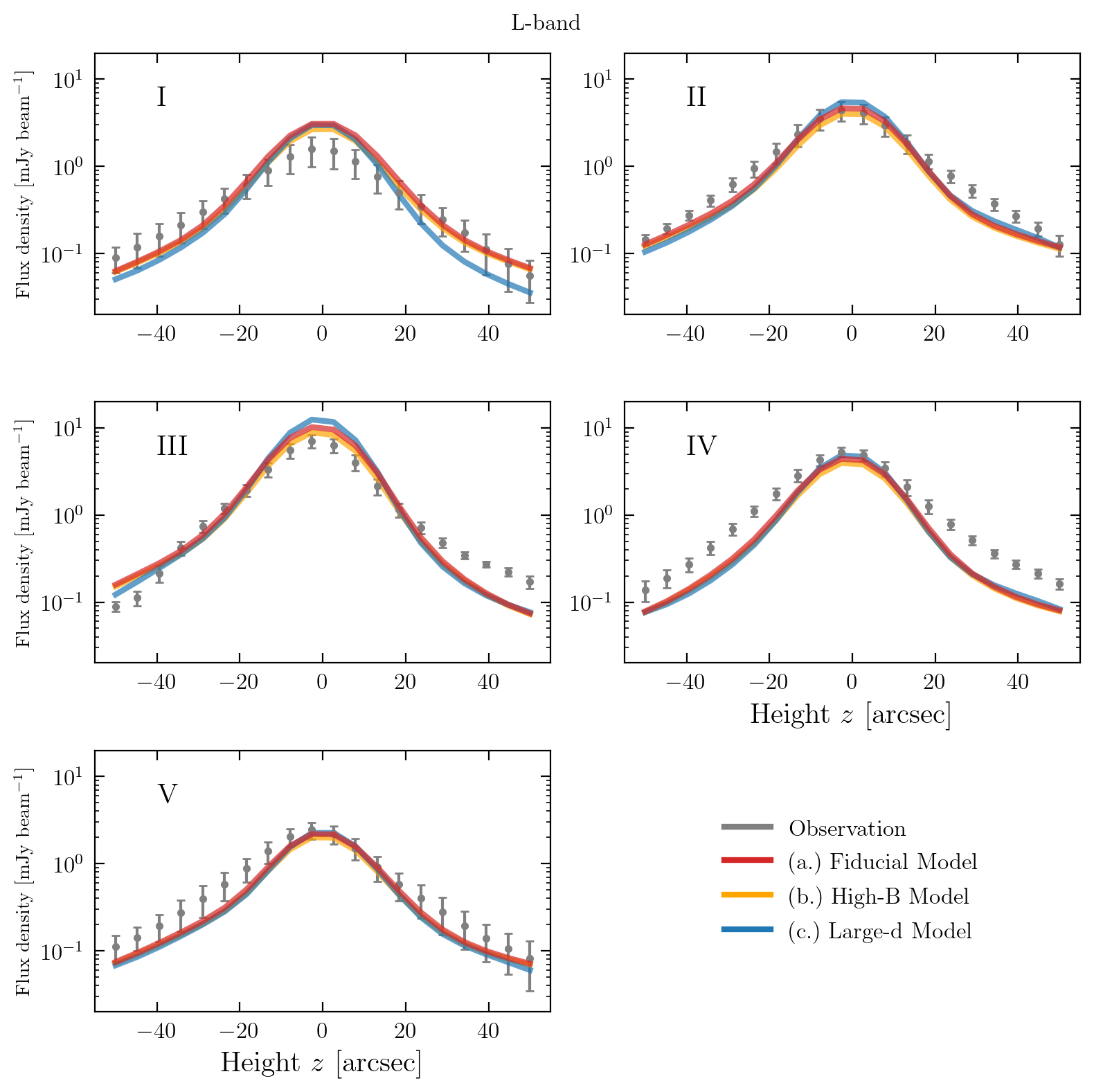}
    \caption{
    Average intensity profile of five different strips at 1.6 GHz. Gray dots present the observation; red line corresponds to our fiducial model; orange line corresponds to our simulation with magnetic field boosted by an additional factor of 2; blue line shows the simulated profile with a larger assumed distance $d=18.88$~Mpc. All profiles vary significantly between different strips.
    }
    \label{fig:strip-Lband}
\end{figure*}
\begin{figure*}
    \centering
    \includegraphics[width=\linewidth]{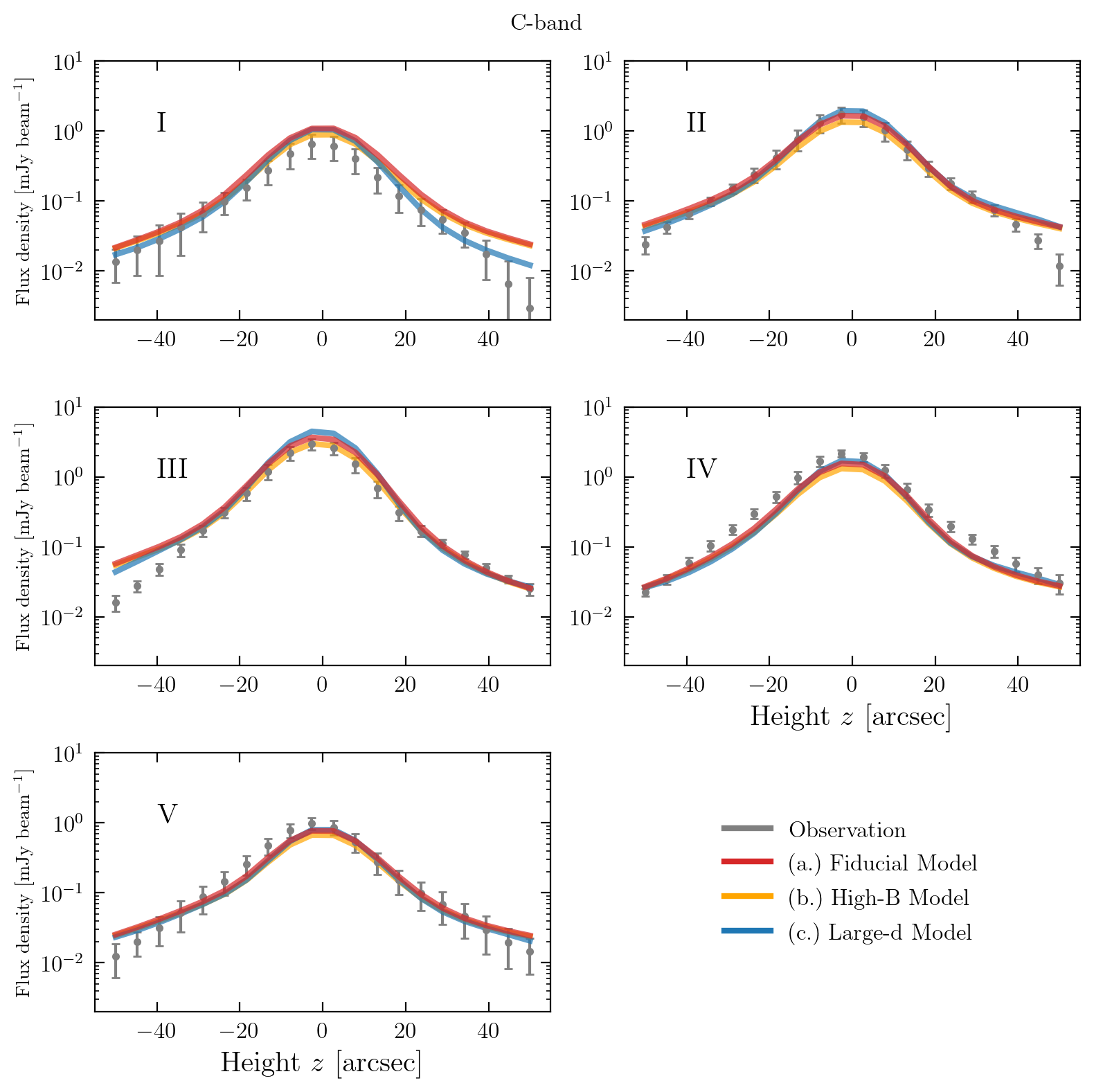}
    \caption{Same as Fig.~\ref{fig:strip-Lband}, but shown at a frequency of 6~GHz.}
    \label{fig:strip-Cband}
\end{figure*}

\bibliography{sample631}{}
\bibliographystyle{aasjournal}


\end{CJK*}
\end{document}